\documentclass[sigconf,nonacm]{acmart}

\RequirePackage{textcase}
\makeatletter
\expandafter\def\expandafter\@secfont\expandafter{\@secfont\MakeTextUppercase}
\makeatother

\usepackage{xspace}
\usepackage{balance}
\usepackage{subcaption}
\usepackage{enumitem}
\usepackage{tikz}
\usepackage{xcolor}
\usepackage{booktabs}
\usepackage{tabularx}
\usepackage{array}

\newcolumntype{Y}{>{\raggedright\arraybackslash}X}
\setcitestyle{numbers,sort,nocompress}
\newcommand{\circleNum}[2][1.2em]{%
  \tikz[baseline=(char.base)]{
    \node[shape=circle, fill=black, text=white, inner sep=0pt,
      minimum size=#1, font=\bfseries] (char) {#2};
  }%
}
\newcommand{\circleOne}[2][1.2em]{%
  \tikz[baseline=(char.base)]{
    \node[shape=circle, draw, inner sep=0pt,
      minimum size=#1] (char) {#2};
  }%
}
\newcommand{\sysname}{\textsc{PrairieFire}\xspace}

\begin{document}

\title{\vspace{3mm}\textsc{PrairieFire}: Igniting High-Throughput Updates in Large-Scale  Approximate Nearest Neighbor Search \vspace{2mm}}
\author{Yuchen Huang}
\affiliation{\institution{East China Normal University}}
\affiliation{\institution{Xiaohongshu Inc (RedNote)}}
\email{ychuang@stu.ecnu.edu.cn}

\author{Baiteng Ma}
\affiliation{\institution{East China Normal University}}
\affiliation{\institution{Xiaohongshu Inc (RedNote)}}
\email{btma@stu.ecnu.edu.cn}

\author{Yiping Sun}
\affiliation{\institution{Xiaohongshu Inc (RedNote)}}
\email{sunyiping@xiaohongshu.com}

\author{Xiao Chen}
\affiliation{\institution{Xiaohongshu Inc (RedNote)}}
\email{chenxiao4@xiaohongshu.com}

\author{Rui Ma}
\affiliation{\institution{Xiaohongshu Inc (RedNote)}}
\email{marui4@xiaohongshu.com}

\author{Yue Chen}
\affiliation{\institution{Xiaohongshu Inc (RedNote)}}
\email{chenyue14@xiaohongshu.com}

\author{Jinwei Zhang}
\affiliation{\institution{Xiaohongshu Inc (RedNote)}}
\email{zhangjinwei1@xiaohongshu.com}

\author{Xiaocheng Zhong}
\affiliation{\institution{Xiaohongshu Inc (RedNote)}}
\email{zhongxiaocheng@xiaohongshu.com}

\author{Zhiyong Wang}
\affiliation{\institution{Xiaohongshu Inc (RedNote)}}
\email{sunzhenghuai@xiaohongshu.com}

\author{Yao Hu}
\affiliation{\institution{Xiaohongshu Inc (RedNote)}}
\email{xiahou@xiaohongshu.com}

\author{Erci Xu}
\affiliation{\institution{Shanghai Jiao Tong University}}
\email{jostep90@gmail.com}

\author{Chuliang Weng}
\affiliation{\institution{East China Normal University}}
\email{clweng@dase.ecnu.edu.cn}

\renewcommand{\shortauthors}{Yuchen Huang et al.}

\makeatletter
\let\paper@original@mkauthors\@mkauthors@iii
\renewcommand{\@mkauthors@iii}{%
  \if@ACM@anonymous
    \paper@original@mkauthors
  \else
    \global\setbox\mktitle@bx=\vbox{%
      \unvbox\mktitle@bx\par\medskip
      \centering\normalfont
      {\Large
        Yuchen Huang\textsuperscript{\dag\ddag},
        Baiteng Ma\textsuperscript{\dag\ddag},
        Yiping Sun\textsuperscript{\ddag},
        Xiao Chen\textsuperscript{\ddag},
        Rui Ma\textsuperscript{\ddag},
        Yue Chen\textsuperscript{\ddag},
        Jinwei Zhang\textsuperscript{\ddag}\par
        Xiaocheng Zhong\textsuperscript{\ddag},
        Zhiyong Wang\textsuperscript{\ddag},
        Yao Hu\textsuperscript{\ddag},
        Erci Xu\textsuperscript{\S},
        Chuliang Weng\textsuperscript{\dag}\par
      }
      {\large\itshape
        \textsuperscript{\dag}East China Normal University,\quad
        \textsuperscript{\ddag}Xiaohongshu Inc (RedNote),\quad
        \textsuperscript{\S}Shanghai Jiao Tong University\par
      }
      {\normalsize
        \textsuperscript{\dag}\{ychuang, btma\}@stu.ecnu.edu.cn,\quad
        \textsuperscript{\ddag}\{sunyiping, chenxiao4, marui4, chenyue14, zhangjinwei1\}@xiaohongshu.com\par
        \textsuperscript{\ddag}\{zhongxiaocheng, sunzhenghuai, xiahou\}@xiaohongshu.com,\quad
        \textsuperscript{\S}jostep90@gmail.com,\quad
        \textsuperscript{\dag}clweng@dase.ecnu.edu.cn\par
      }
      \bigskip
      \vspace{7mm}
    }%
  \fi
}
\makeatother

\begin{abstract}
Arrays of modern NVMe SSDs are approaching memory in terms of bandwidth in a single server while offering much lower cost per gigabyte. Hence, approximate nearest neighbor search (ANNS) for billion-scale vectors in applications such as recommendation, search, and retrieval-augmented generation increasingly runs on SSD-based systems. Unfortunately,
due to the limited update throughput of existing SSD-based solutions, the indexed corpus cannot keep pace with continuous modifications to data and embedding models, thus hindering broader deployment across more scenarios and larger scales.

The bottleneck lies in CPU-intensive update paths, where foreground insertions incur frequent SSD accesses and background reassignments repeatedly relocate each vector, making high update throughput difficult to achieve even by exhausting available hardware.
In this paper, using buffered updates and batched local rebuilds, we develop \sysname with cascaded adjustments, concurrency control, and CPU-GPU-SSD cooperation. These co-designs
avoid frequent SSD accesses in foreground updates and heavy background reassignments, achieving high update throughput while maintaining excellent search performance.

Evaluations show that, compared with state-of-the-art systems, \sysname{} can sustain $\sim$3-35$\times$ update throughput while maintaining $\sim$3-5$\times$ search throughput simultaneously. It also achieves $\sim$1.5-8.7$\times$ cost efficiency. \sysname has been deployed in RedNote's production environments for two months, improving business metrics (e.g., click-through rate) by higher freshness without frequent index rebuilds.

\end{abstract}

\maketitle

\pagestyle{plain}
\thispagestyle{plain}

\section{Introduction}
\setlength{\parskip}{0pt plus 1pt}
Approximate nearest neighbor search (ANNS) over embedding vectors underpins online services such as recommendation, search engines, advertising, and retrieval-augmented generation~\cite{FreshRec,CHAVS,JDPlatform,Chameleon,RAGO}. To reduce storage cost over billions of vectors, recent systems try to place raw vectors as cluster lists or as graph nodes with neighboring edges from DRAM onto NVMe SSDs~\cite{DiskANN,SPANN,PipeANN,Starling,Helmsman}. The corpus of vectors, however, is anything but static. In upstream applications, items are continuously created and retired, and the underlying embedding models are continuously adjusted and synchronized at an hourly or finer cadence to track shifting user behavior for better business outcomes~\cite{FreshRec,Monolith}. A practical ANNS service must therefore find top-$k$ results against the latest valid data while continuously absorbing a stream of inserts and deletes arising from both dynamic items~\cite{SPFresh,OdinANN} and continual learning of embedding models~\cite{JLDRM,QDC}.

To support updates, based on existing static ANNS~\cite{DiskANN,SPANN}, two notable solutions have been proposed (i.e., OdinANN~\cite{OdinANN} and SPFresh~\cite{SPFresh}). OdinANN follows the graph  structure~\cite{DiskANN} and introduces direct modifications to avoid the inefficiency of periodic merging (e.g., 1.5-2$\times$ search latency and $\sim$3$\times$ peak memory usage during merging~\cite{FreshDiskANN,TopologyAwareLocalizedUpdate}). SPFresh adopts a partition-based structure, which routes vectors to partitions through an in-memory centroid index and maintains partitions' lists on SSDs. To address the degraded index quality under continuous updates, SPFresh introduces a local rebuild mechanism that splits or merges partitions and reassigns vectors to their nearest partitions.

However, when extending SSD-based ANNS to broader large-scale scenarios, their update performance becomes a limitation. Recent industrial practice points out that, to track user behavior in real time, the continual learning~\cite{JLDRM,QDC,FreshRec,Monolith} of embedding models calls for high-throughput updates. For a 0.1B-scale index in recommendation systems, maintaining freshness demands an update throughput of up to 30K queries per second (QPS) to match~\cite{Helmsman}. This requirement is comparable to the search throughput of online services, making updates the other first-class performance target. But even if we deploy an SSD array with >100GB/s bandwidth and exhaust all 64 CPU cores on update-only operations, OdinANN and SPFresh can only sustain about 6\% to 40\% of the target.

Graph-based OdinANN achieves only 2--3 KQPS because each update searches $\sim$100 neighbors before modifying links, incurring $\sim$4--7 ms due to inherent multi-hop SSD accesses~\cite{TridentANN,Starling,Helmsman}. Partition-based SPFresh routes inserts through in-memory centroids ($<$1 ms), but rewriting SSD posting lists per insertion in foreground still incurs unacceptable CPU consumptions. Background per-vector relocations after splits or merges require extra CPU cores to deal with centroid locations and partition updates. Even with 64 CPU cores for update-only requests, it can only reach $\sim$12 KQPS.

Hence, we propose to use buffered updates and batched reclustering for rebuilds, instead of immediate flushes and asynchronous reassignments in SPFresh. First, buffered updates can amortize the overhead of foreground SSD round trips, where multiple updates to the same cluster are accumulated in memory and flushed to SSD as a batch. Second, we can recluster an imbalanced region, replacing the steady drip of per-vector reassignments with a single batch local rebuild. However, it is non-trivial to apply these ideas. There are several conflicts and challenges.

\setlength{\parskip}{0pt}
\textbf{Conflicts.} For high-performance search, SPFresh takes fine-grained partitioning
and replicated placement. However, these designs leave no memory budget for per-list buffers, and make the buffering less effective. SPFresh creates up to 16\% as many partitions as vectors, so even small per-partition buffers (e.g., 8 or 16 vectors) can raise memory footprint to 2--3$\times$. And a single new vector can trigger up to 8$\times$ insertions, which again cause frequent SSD reads/writes even with buffered updates.

\textbf{Challenges.} Hence, we return to normal-scale partitions (e.g., 1-2\% of the dataset size) and single-copy placement, but this burdens search. In detail, it can suffer more from distribution drift, since updates are amortized across fewer partitions, each partition experiences more modifications, thus reducing the representativeness of static centroids. Without replications, more clusters need to be searched for each query simultaneously to keep accuracy. To address these, we need dynamic centroid adjustment to track updates, high concurrency and better pipelines to support search performance.
Moreover, we use reclustering to replace reassignments after splits or merges, but reclustering incurs a higher computational burden and requires equal or more cores than background reassignments, making reclustering inefficient.

In this paper, we propose \sysname{} with the following key techniques:
\begin{itemize}[leftmargin=*, nosep]
\item Based on above analysis, \sysname organizes a three-level partition-based index and supports cascaded updates on centroids, where insertions and deletions of vectors can be buffered and propagated to upper-level centroids, thus avoiding frequent SSD round trips and preserving centroid representativeness.
\item Based on the index structure, we develop fine-grained concurrency control to support high concurrency of operations, including insertions, deletions, searches and local rebuilds.
\item We build the co-designed workflow on GPUs, CPUs and high-bandwidth SSDs to accelerate centroid routing, partition scanning, and computing-intensive clustering-based rebuilds.
\end{itemize}

We compare \sysname{} with state-of-the-art ANNS that support updates on large-scale workloads. Comprehensive evaluations show that \sysname{} can sustain $\sim$3-35$\times$ update throughput while maintaining $\sim$3-5$\times$ search throughput simultaneously. In addition, during the whole process, our search accuracy is comparable to or higher than baselines. At last, we evaluate the cost and energy efficiency of \sysname{}, showing up to 1.5-8.7$\times$ hardware efficiency of baselines in terms of device acquisition, power consumption, and storage footprint. \sysname has been running reliably in RedNote's services for two months, reducing index rebuild frequency by over 95\% across the dozens of indexes it hosts and improving all business metrics (e.g., click-through rate and content impressions).

\section{Background}

\subsection{Freshness Challenges of Large-Scale ANNS}

ANNS has become a fundamental infrastructure component~\cite{RaBitQ,HAKES,HDANNS} for large-scale search, recommendation, and advertising services over embedding collections that often contain billions of vectors. To reduce CapEx and OpEx at this scale, recent ANNS systems increasingly rely on SSD-based deployments~\cite{DiskANN,SPANN,PipeANN,Starling}.
Since modern SSD arrays can  provide sufficient bandwidth approaching that of DRAM~\cite{whatNVMeVLDB,NVMeSSDBandwidth}, we have deployed SSD-based Helmsman~\cite{Helmsman} at scale for online services in RedNote (a.k.a. Xiaohongshu)~\cite{RedNote}. Under strict 5-10ms search latency service level agreement (SLA), it only consumes dozens of multi-SSD servers to handle vector search services that previously required clusters equipped with tens of thousands of CPU cores and PB-level DRAM.

However, if it can only support search on static collections, using SSD-based storage~\cite{Helmsman} to offload memory pressure alone would be insufficient. In practice, vector collections are highly dynamic and require continuous updates. Changes to products, articles, and embeddings must be reflected promptly to avoid stale or invalid results~\cite{Milvus,FreshDiskANN,Neos}. In the search services, daily  full refreshes are common to keep the index up-to-date. Moreover, recommendation systems frequently retrain and deploy embedding models at hourly or finer intervals to capture evolving user behavior~\cite{FreshRec,Monolith}. These model changes can modify vector representations, requiring corresponding index updates~\cite{JLDRM,QDC}.

\subsection{Operations for Dynamic Workloads}
To make the dynamic vector changes timely visible in
these services, three
operations~\cite{OdinANN,SPFresh,TopologyAwareLocalizedUpdate,SurveyVDBMS} are required:
\begin{itemize}[leftmargin=*, nosep]
    \item \textbf{\textit{Search (Top-$k$, Vector)}} takes a query vector as
    input and returns the top-$k$ approximate nearest neighbors from the current
    valid collection. In dynamic ANNS, search should be performed over the
    real-time state of the collection, reflecting preceding insertions and
    deletions.

    \item \textbf{\textit{Insert (ID, Vector)}} inserts a new vector into the
    collection, where \textit{ID} denotes the unique identifier of the
    vector and \textit{Vector} denotes its embedding representation. After
    insertion, the new vector should be included to subsequent searches and potential results.

    \item \textbf{\textit{Delete (ID)}} removes an existing vector from the
    collection according to its \textit{ID}. After deletion, the removed vector
    would no longer appear in subsequent search results.
\end{itemize}

\subsection{Existing Solutions}
\noindent\textbf{Graph-based solution.} Recent works~\cite{FreshDiskANN,TopologyAwareLocalizedUpdate,OdinANN}
demonstrate the capability to support
updatable ANNS by online maintaining the SSD-resident graph structure~\cite{DiskANN}.

Specifically, FreshDiskANN and Greator~\cite{FreshDiskANN,TopologyAwareLocalizedUpdate} buffer newly inserted vectors in
an in-memory graph and periodically merge them in batches into the global on-disk
graph index.  However, this mechanism can incur unacceptable memory overhead
and suboptimal search performance during graph merging (e.g., 32-GB peak memory usage and 1.5-2$\times$ search latency when merging 6M vectors on 60-GB index of 100M vectors). To
address these limitations, OdinANN~\cite{OdinANN} adopts direct insertion, which
searches neighboring nodes for each insertion, updates related edges, creates a
new node, and directly flushes the changes to the NVMe SSD. By avoiding periodic
merging and optimizing immediate maintenance, OdinANN provides higher and
more stable performance of updates and search with lower memory usage.

\noindent\textbf{Partition-based solution.} Following the partition-based structure of SPANN~\cite{SPANN},
SPFresh~\cite{SPFresh} stores
partition-level metadata in memory (e.g., centroid index,
concurrency locks, and cluster sizes) and stores vector partitions as
posting lists on the NVMe SSD. It supports streaming
updates with a lightweight incremental rebalancing protocol, namely
\textit{LIRE}.

Under continuous updates, partitions can become imbalanced.
With \textit{LIRE}, SPFresh locally splits oversized partitions, merges undersized partitions, and reassigns vectors to the nearest partitions, thereby adapting the index without full rebuilding. As inherited from SPANN, SPFresh follows search-optimized layout through aggressive fine-grained clustering (i.e., creating
centroids whose number reaches around 12--16\% of the total vector count) and
redundantly inserts vectors into multiple clusters' posting lists on the NVMe
SSD~\cite{SPANN,SPFresh,RAIRS}.

\section{Motivation}
\label{sec:motivation}
\subsection{Insufficient Performance}

 \begin{figure}[t]
	\centering
	\begin{subfigure}[t]{0.49\linewidth}
		\includegraphics[width=\linewidth]{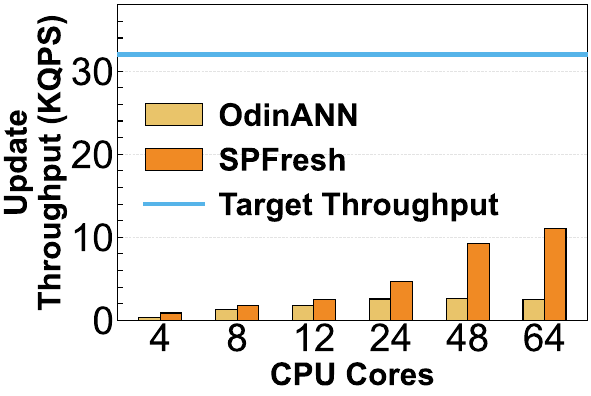}
				\vspace{-7mm}
		\caption{\centering\textbf{Increasing CPU cores with 1 PCIe-Gen5 NVMe SSD.}}
		\label{fig:mot31-core}
		\vspace{-4mm}
	\end{subfigure}
	\centering
	\begin{subfigure}[t]{0.49\linewidth}
		\includegraphics[width=\linewidth]{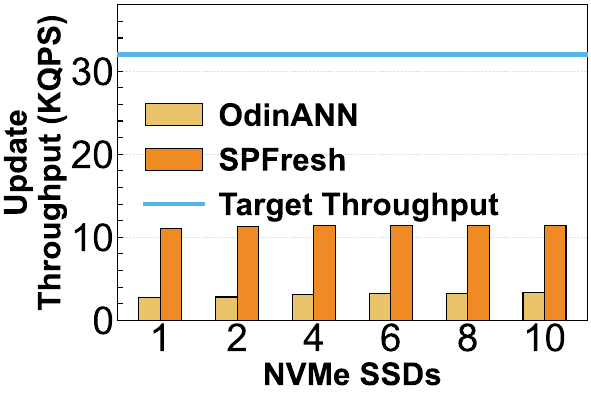}
			\vspace{-7mm}
		\caption{\centering\textbf{Increasing SSDs with 64 cores enabled.}}
		\label{fig:mot31-ssd}
		\vspace{-4mm}
	\end{subfigure}
	\centering
	\caption{Based on SIFT100M, scaling throughput of updates (insert/delete = 1/1) with CPUs (a) and SSDs (b). However, scaling both resources fails to achieve the target 30K QPS.}
	\label{fig:mot31-scaling}
	\vspace{-5mm}
\end{figure}

Since items and embedding models are continuously generated and modified to
improve user experience and revenue-related metrics~\cite{JDPlatform,FreshRec},
we require real-time index updates to ensure embedding freshness.
To replace periodic full-index rebuilding, online updates must sustain the
required refresh rate, estimated as the index cardinality divided by the
full-refresh interval~\cite{Helmsman}.
Representative workloads~\cite{RTAMS-GANNS,Monolith}, rebuilding a
billion-vector index every 10 hours or a 100-million-vector index every
hour, thus require $\sim$30K QPS of updates (1 insertion and 1 deletion per update).

However, \textbf{existing state-of-the-art solutions fall short of production update demands.}
Even under optimistic update-only workloads with concurrent searches
disabled, state-of-the-art systems OdinANN and SPFresh fall far short of this
target. As shown in~\autoref{fig:mot31-core}, even when dedicating all threads to updates
on up to 64 CPU cores, they yield only 2--12K QPS in a single-server setup, at most 40\% of the required
30K QPS.

OdinANN is bottlenecked by its insertion path. During its insertion, each new vector must first search the SSD-resident graph to identify the top-100 (i.e., the number of neighbors to be connected by new edges) nearest existing nodes, after which OdinANN acquires exclusive locks on these nodes to apply the update. These multi-hop SSD accesses from searching existing neighbors introduce most of the overhead (i.e., 4--7 ms, up to 80\% of the total time), while lock contention from nodes' updates further limits concurrency of this graph-based structure.

Partition-based SPFresh delivers better update throughput than graph-based OdinANN because it only searches an in-memory centroid index to locate several nearest clusters for insertion, which is faster (i.e., <1 ms) than OdinANN's on-SSD top-100 graph search. It then locks these clusters and updates their SSD pages with batched I/O. However, it still remains $\sim$60\% below the required throughput when all threads are used for updates.

\subsection{Does Bandwidth Scaling Work?}

\textbf{Scaling SSD bandwidth fails.} To break the bandwidth bottleneck of partition-based solutions, we have deployed multiple SSDs to improve search throughput in industrial services~\cite{Helmsman}, which inspires us to explore whether updates benefit similarly from this increased I/O bandwidth of modern NVMe SSD arrays~\cite{GustANN,FlashANNS,TridentANN,Helmsman}.

As shown in~\autoref{fig:mot31-ssd}, results are disappointing. First, higher bandwidth provides little benefit to OdinANN, whose performance is dominated by SSD latency rather than bandwidth, due to its multi-hop graph-based access pattern~\cite{GustANN}.
Then, we might expect SPFresh to benefit more from increased I/O bandwidth, given its bandwidth-friendly batched I/O pattern and partition-based index structure~\cite{SPFresh,Helmsman}.
Unfortunately, SPFresh also improves hardly in update performance at all, far from the near-linear scaling often observed in the pure search-only workloads~\cite{Helmsman}. Even with 10 Gen5 NVMe SSDs providing almost an order of magnitude more available bandwidth (i.e., $\sim$120 GB/s), update throughput improves by less than 10\%.

 \begin{figure}[t]
	\centering
	\begin{subfigure}[t]{0.49\linewidth}
		\includegraphics[width=\linewidth]{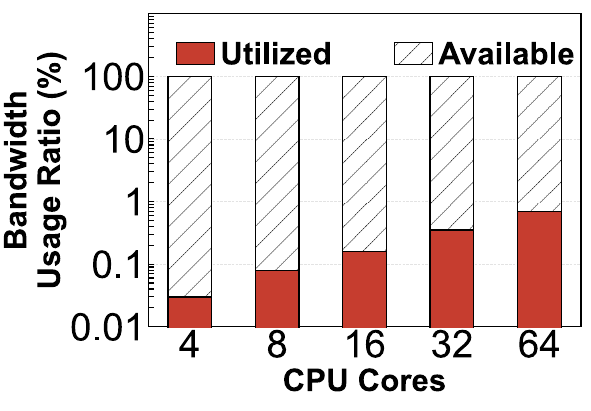}
				\vspace{-7mm}
		\caption{\centering\textbf{Bandwidth utilization with increasing cores.}}
		\label{fig:mot32-bd}
		\vspace{-4mm}
	\end{subfigure}
	\centering
	\begin{subfigure}[t]{0.49\linewidth}
		\includegraphics[width=\linewidth]{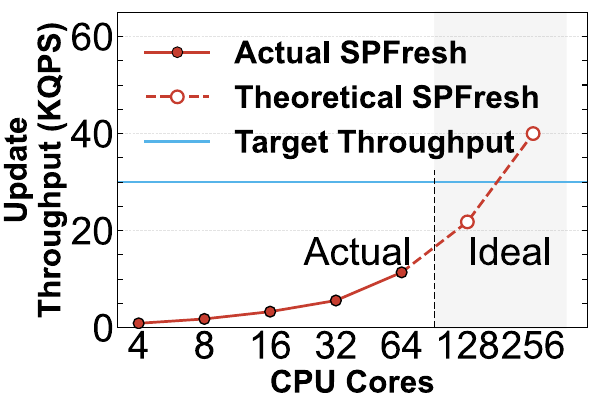}
		\vspace{-7mm}
		\caption{\centering\textbf{Cores required to achieve target throughput ideally.}}
		\label{fig:mot32-ideal}
		\vspace{-4mm}
	\end{subfigure}
	\centering
	\caption{Increasing CPUs under suitable thread allocations~\cite{SPFresh} (i.e., 1 thread for delete, others for insert and background as an approximate ratio of 2:1). We compare the actual bandwidth usage with the available value (a), and show theoretical cores required to achieve the target throughput (b).}
	\label{fig:mot32-why}
	\vspace{-5mm}
\end{figure}

To understand the reason, we analyze the bandwidth utilization and the achievable throughput as scaling the update concurrency by increasing cores, as shown in~\autoref{fig:mot32-why}.

\textbf{For SPFresh updates, bandwidth is not the bottleneck, but the high CPU core consumption is.} First, we measure the actual bandwidth usage of 10 PCIe-Gen5 SSDs for SPFresh as we increase update concurrency.
As shown in~\autoref{fig:mot32-bd}, even with 10 SSDs providing $\sim$120 GB/s of bandwidth, the bandwidth usage remains below 1 GB/s, indicating that the available bandwidth is not the bottleneck. Hence, we further estimate how many CPU cores for concurrent updates would be required to achieve the target.

Unfortunately, as shown in ~\autoref{fig:mot32-ideal}, though partition-based SPFresh already reduces modification-selection time (i.e., only locating the relevant clusters in memory but not searching neighbors on SSDs as OdinANN), it still requires 128--256 cores theoretically for the target throughput even without reserving CPU cores for search requests, which is far beyond the available hardware resources in real-world industrial deployments.

\subsection{Where Does CPU Cost Come From?}
To address the impractical consumption of CPU cores required for the target update throughput, we first look back at the update procedures in SPFresh, and then analyze where the CPU consumption comes from.

\begin{figure}[t]
	\centering
	\includegraphics[width=0.9\columnwidth]{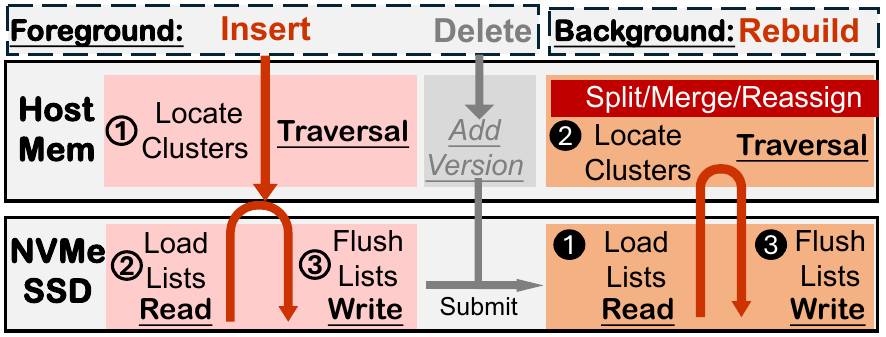}
	 \vspace{-3mm}
	\caption{Update workflow in SPFresh. Main CPU overheads are from foreground insertion and background rebuilds.}
	\label{fig:mot33-spfresh}
	\vspace{-7mm}
\end{figure}

\textbf{Procedure of updates.} In SPFresh, update threads consist of both foreground threads and background threads.
As shown in~\autoref{fig:mot33-spfresh}, SPFresh supports \textit{Insert} and \textit{Delete} operations in foreground threads, and performs \textit{rebuild} operations to maintain the index quality in background threads.

For \textit{Insert}, \circleOne[0.8em]{1} it first locates relevant clusters for the new vector by searching the in-memory centroid graph via node traversal. Then, it updates these clusters' SSD lists with read-modify-write round-trip I/Os (i.e., \circleOne[0.8em]{2} reads for loading and \circleOne[0.8em]{3} following writes for flushing modifications). \textit{Delete} is lightweight because it only adds the version value of the deleted vector's \textit{ID} in the in-memory version map. Hence, most of the foreground update threads are allocated for \textit{Insert} (e.g., $\sim$40 cores, two thirds of the total in update-only stress tests) and only 1 core is enough for \textit{Delete} of SPFresh~\cite{SPFresh}.

For \textit{rebuild}, when the size of clusters exceeds a threshold (e.g., >128) or undersized (e.g., <10), the rebuilds are triggered to split oversized clusters into two smaller ones or directly append small clusters into a larger one. Then, to address the possible violation of nearest partition assignments (NPA), it reassigns vectors from these rebuilt clusters. In this process, it would first \circleNum[0.8em]{1} read vector lists of these clusters from SSDs, then \circleNum[0.8em]{2} check whether vectors need reassignment, relocate the nearest clusters for vectors potentially violating NPA, and \circleNum[0.8em]{3} finally write them back to SSDs. Due to the expensive reassignments~\cite{SPFresh}, the rebuilds are performed with the other one third of the total cores in the background.

\begin{figure}[t]
	\centering
	\includegraphics[width=0.82\columnwidth]{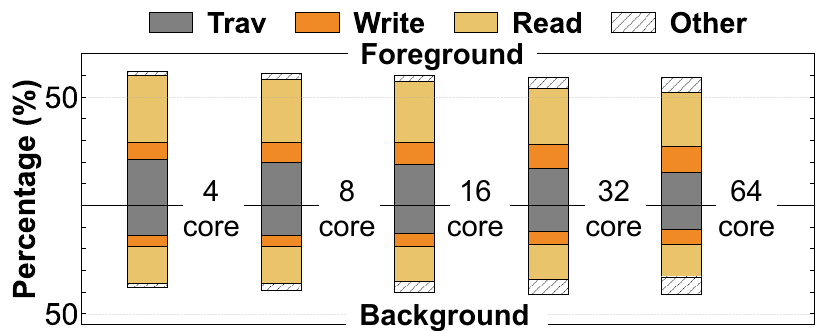}
	 \vspace{-3mm}
	\caption{Breakdown of CPU cycles. \textbf{Trav}: locate clusters over the centroid graph; \textbf{Write}: flush lists to SSDs; \textbf{Read}: load lists from SSDs. \textbf{Other}: remaining activities (e.g., lock spinning).}
	\label{fig:mot33-insert}
	\vspace{-5mm}
\end{figure}

\textbf{Consumption breakdown.}
Based on Perf~\cite{PerfLinux}, we analyze the CPU cycles spent on foreground and background, as shown in~\autoref{fig:mot33-insert}. In update-only stress tests with 4--64 threads, foreground updates account for 59--62\% of the total cycles, while background rebuilds contribute the remaining 38--41\%.
Both paths repeatedly perform per-vector centroid searches
and posting-list updates by traversing in-memory graph and reading/writing their lists on SSDs. Specifically, centroid search by graph traversal accounts for around 30--35\% of cycles, while round-trip read-modify-write updates dominate the remaining cycles.

Foreground insertions trigger SSD reads and writes for every new vector. As a result, updates generate frequent SSD accesses, consuming huge CPU resources in I/O processing (e.g., submitting commands, polling for completions) and clusters updating (e.g., modifying lists and metadata).
And under continuous updates, clusters can become imbalanced, triggering background rebuilds for splitting or merging clusters to maintain the index quality. Then, vectors can violate the NPA after splitting/merging imbalanced clusters, necessitating background reassignments. The per-vector reassignment of vectors further increases CPU cycle consumption in background by triggering centroid searches and rewriting on-SSD lists for the vectors affected.

\section{Opportunities and Challenges}
\subsection{Buffered Updates and Direct Rebuilds}
Based on these observations, we identify two intuitive opportunities for improving index maintenance efficiency and saving CPU cycles to achieve higher update throughput.

\textbf{Using buffered updates to amortize frequent round-trip I/Os of per-vector read-modify-write operations in the foreground threads.}
We can buffer new insertions and deletions in memory for each cluster instead of being immediately written to SSDs. Thus, by accumulating them, we can minimize every cluster's lists read-modify-write I/Os and flush them to the corresponding SSD pages in a merged batch. As a result, foreground CPU cycles spent on read/write operations from per-vector insertions can be substantially reduced.

\textbf{Using local reclustering-based rebuilds to avoid per-vector relocation of reassignment in the background threads.}
When an imbalanced cluster is split or merged, we can directly fetch its nearby clusters and recluster them to generate new centroids and posting lists. This preserves partition quality by reconstructing neighboring clusters jointly, rather than repeatedly relocating individual vectors~\cite{Quake,IncrementalIVF}. As a result, this generalized \textit{LIRE} can replace per-vector centroid searches and I/Os during relocations with a single centroid search for the merged/split cluster's centroid and batched clusters refresh.

\label{sec:clgAndCns}
\subsection{Conflicts and Challenges Arising}
\begin{figure}[t]
	\centering
	\begin{subfigure}[t]{0.32\linewidth}
		\includegraphics[width=\linewidth]{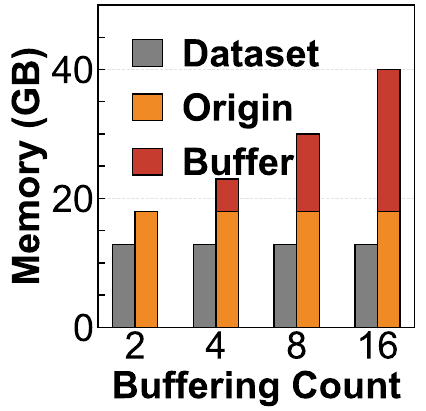}
			\vspace{-5mm}
		\caption{\centering\textbf{Memory\\bloating.}}
		\label{fig:cha41-mb1}
		\vspace{-4mm}
	\end{subfigure}
	\centering
	\begin{subfigure}[t]{0.32\linewidth}
		\includegraphics[width=\linewidth]{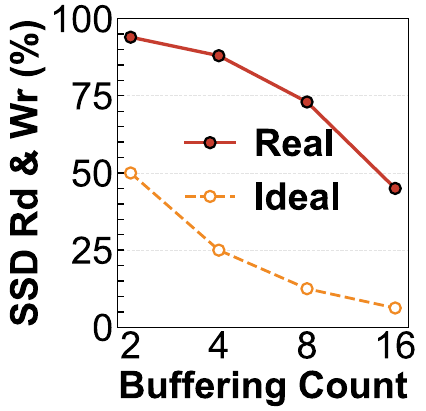}
			\vspace{-5mm}
		\caption{\centering\textbf{Pseudo\\buffering.}}
		\label{fig:cha41-rp1}
		\vspace{-4mm}
	\end{subfigure}
	\centering
    	\begin{subfigure}[t]{0.32\linewidth}
		\includegraphics[width=\linewidth]{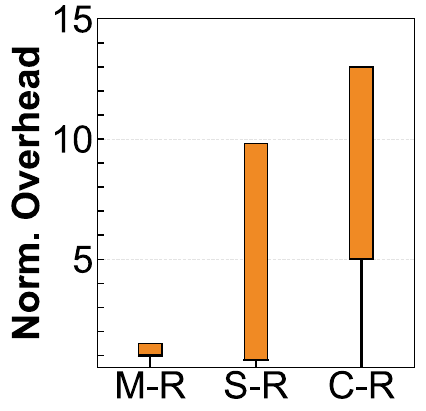}
			\vspace{-5mm}
		\caption{\centering\textbf{Computation\\burden.}}
		\label{fig:cha41-cb1}
		\vspace{-4mm}
	\end{subfigure}
	\caption{We show memory usage comparisons (a) and SSD I/O trigger ratios of insertion (b) across various buffer sizes. (c) shows the CPU time needed by merge reassignment (M-R), split reassignment (S-R), and direct clustering rebuilds (C-R).}
	\label{fig:cha41-1}
	\vspace{-2mm}
\end{figure}

\begin{figure}[t]
	\centering
	\begin{subfigure}[t]{0.32\linewidth}
		\includegraphics[width=\linewidth]{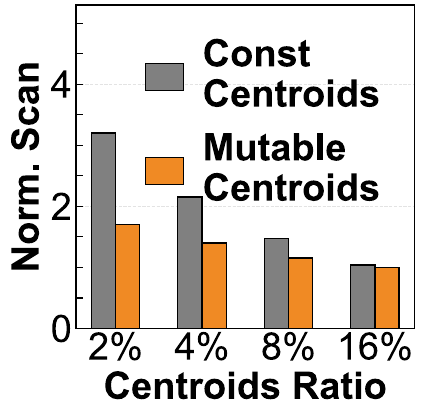}
				\vspace{-5.5mm}
		\caption{\centering\textbf{Amplified scan.}}
		\label{fig:cha41-mb2}
		\vspace{-3mm}
	\end{subfigure}
	\centering
	\begin{subfigure}[t]{0.32\linewidth}
		\includegraphics[width=\linewidth]{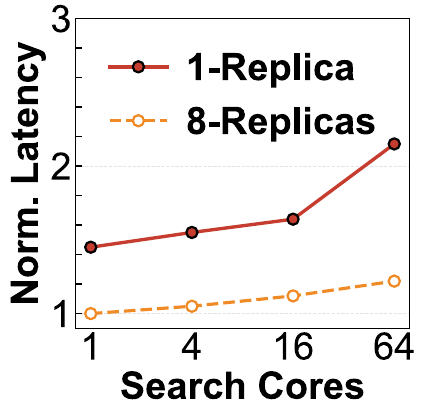}
		\vspace{-5.5mm}
		\caption{\centering\textbf{Higher latency.}}
		\label{fig:cha41-rp2}
		\vspace{-3mm}
	\end{subfigure}
	\centering
    	\begin{subfigure}[t]{0.32\linewidth}
		\includegraphics[width=\linewidth]{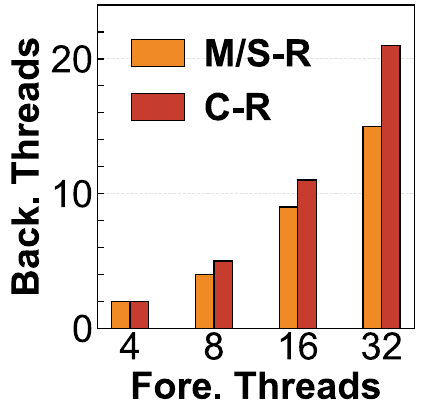}
				\vspace{-5.5mm}
		\caption{\centering\textbf{More cores.}}
		\label{fig:cha41-cb2}
		\vspace{-4.5mm}
	\end{subfigure}
	\caption{(a) shows normalized cluster scanned for recall@10=90\% when reducing centroids, with/without centroid modification. (b) shows search latency before/after removing replicas. (c) shows background cores of rebalancing required to match varying foreground cores of insertion with merge-split reassignment (M/S-R) or direct clustering rebuilds (C-R).}
	\label{fig:cha41-2}
	\vspace{-2mm}
\end{figure}

However, directly applying these ideas in SPFresh does not work well. These two intuitive methods would conflict with current search-oriented design choices and limited hardware capabilities. We discuss three conflicts (\autoref{fig:cha41-1}) and their corresponding challenges (\autoref{fig:cha41-2}) that prevent buffering and reclustering from being effective in practice, as follows.

\textbf{Memory bloating.}
The first conflict arises from SPFresh's fine-grained partitioning and buffering. Following SPANN, SPFresh creates up to 16\% as many clusters as the total number of vectors to improve search efficiency~\cite{SPANN}. However, maintaining per-list buffers for merging each cluster's SSD flushes further enlarges its already high memory footprint. As shown in~\autoref{fig:cha41-mb1}, SPFresh (i.e., Origin) already consumes about $1.5\times$ the raw dataset size. On the $\sim$13GB SIFT100M dataset, increasing the per-list buffer count from 2 to 16 adds extra 5-22 GB, and at the count of 16, the total memory consumption approaches $3\times$ the dataset size.

One straightforward solution is to reduce the number of clusters to a normal IVF scale~\cite{FAISS}, making memory usage more affordable. However, fewer clusters would absorb more member changes per cluster under the same update rate, making fixed centroids quickly become outdated.
Hence, fixed centroids can consequently become poorly representative of their evolving members, impairing routing accuracy. As shown in~\autoref{fig:cha41-mb2}, after 100M updates, fewer centroids require scanning more clusters to maintain recall, especially when centroids remain unchanged. With a normal-scale centroid count (i.e., $\sim$ 2\% of the vector count, with $\sim$50--100 members per cluster), the scan overhead increases by up to $\sim 3\times$.

\textbf{Pseudo buffering.}
The second issue comes from SPFresh's redundant placement. To mitigate boundary-induced recall loss, SPFresh follows SPANN, using RNG checks~\cite{RNG} to place each vector into up to 8 clusters. This largely defeats buffered insertion: ideally, a buffer size of $B$ should reduce the SSD-access probability to about $1/B$ per insertion. However, as shown in~\autoref{fig:cha41-rp1}, redundancy greatly raises this probability of triggering SSD accesses. With buffering 2 vectors for merging I/Os of each partition, replicated placements still trigger SSD access with nearly 100\% probability; even with a buffer budget of 16 vectors per partition, the probability of round-trip SSD access remains around 40\%.

Therefore, to make buffering updates effective, we argue that redundant placement can be removed. However, this shifts the cost to search. As shown in~\autoref{fig:cha41-rp2}, using single-thread 8-replica search as the baseline, we compare search latency with and without replicas under increasing search concurrency. Without redundancy, search faces higher latency due to reading SSDs and computing distances for more cluster lists; when concurrent search threads reach 64 CPU cores, search latency further increases sharply due to intensified hardware contention on concurrent CPU threads and memory bandwidth of DRAM~\cite{Helmsman}.

\textbf{Computational burden.}
The third issue comes from computing burden of reclustering-based local rebuilds. While the batched clustering avoids per-vector centroid searches and frequent updates on-SSD posting lists for all vectors affected by imbalanced partitions after splitting or merging, it introduces clustering that has heavier computing consumption than centroid search during background reassignments.

As shown in~\autoref{fig:cha41-cb1}, using CPU cycles of relocating all vectors after merging undersized clusters (i.e., list size less than 10~\cite{SPFresh}) as the baseline, we compare per-vector reassignment overhead after splitting/merging with local reclustering 64 nearby clusters~\cite{Quake,SPFresh}. With the common 10-15 Lloyd iterations in K-Means~\cite{FAISS,Lloyd,lloyd1982} to local convergence, direct reclustering can amplify computations. Consequently, as shown in~\autoref{fig:cha41-cb2}, more CPU cores (i.e., C-R) can be required for performing local reclustering in background threads, to match update throughputs under the same CPU cores of foreground threads.

\section{\sysname}
\subsection{Considerations}
Addressing these conflicts and trade-offs, to make buffered update and batched rebuild practical, we have the following considerations.

First, we argue to use \textbf{an index structure with normal partitioning granularity, centroid-adaptive design, and non-redundant placement}. Normal-scale partitioning supports buffered updates without an unacceptable memory footprint. Centroid-adaptive design is for continuously tracking cluster modifications to avoid degraded centroid representativeness. Non-redundant placement eliminates the multiple insertion of the same vector into different clusters, reducing SSD access frequency and avoiding pseudo buffering.

Second, index operations call for an efficient \textbf{concurrency control mechanism} on this update-oriented index structure to proceed with minimal contention, enabling concurrent searches, updates, and index maintenance.

Third, it is practical to introduce \textbf{GPUs as extra computational resources along with efficient pipelines among various hardware} (i.e., GPUs, CPUs, and SSDs) to support heavier search overhead from our update-oriented index structure, and to enable direct clustering-based rebuilds without overwhelming CPUs.

\subsection{Design Overview}
\begin{figure}[t]
	\centering
	\includegraphics[width=0.96\columnwidth]{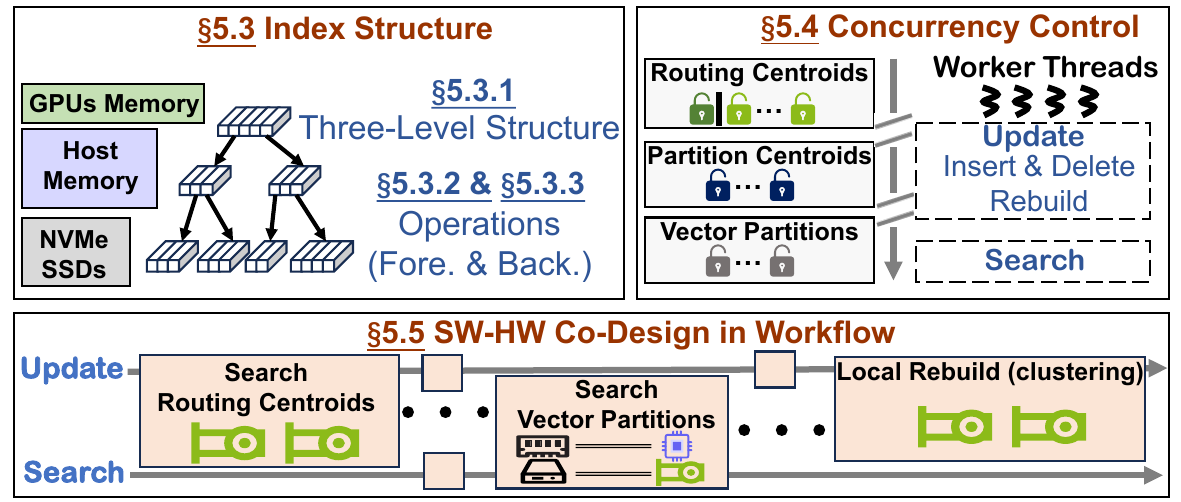}
	 \vspace{-4mm}
	\caption{Design Overview of \sysname.}
	\label{fig:des51-overview}
	\vspace{-6mm}
\end{figure}

Following these considerations, we propose \sysname. \autoref{fig:des51-overview} shows the design overview, including the index structure, concurrency control, and software-hardware co-design pipelines in the update-search workflow.

\textbf{Index structure} (~\S\ref{sec:layout}).
First, with normal-granularity partitioning for affordable buffering, \sysname organizes a three-level partition-based index~\cite{Quake,FAISS} across SSDs, memory, and GPUs. From bottom to top, it consists of vector partitions as level-2, partition centroids as level-1, and routing centroids as level-0 (\S\ref{sec:three-level layout}). Buffering updates and search operations are managed across these levels simultaneously (\S\ref{sec:operation paths}), then local rebuild and cascaded centroid-adaptation are applied to maintain index quality (\S\ref{sec:Cascaded Adjustment}).

\textbf{Concurrency control} (~\S\ref{sec:concurrency}).
To support concurrent online operations including searches, inserts, deletes, and background local rebuilds, we design a granularity-decoupled locking scheme to efficiently reduce concurrency contention. It is designed to narrow the scope of locking by decoupling different levels of the hierarchical partition-based index. Then, we use the locking-validation-execution pattern to ensure consistency, while preventing upper-level locks from unnecessarily blocking operations on unrelated lower-level objects in the index.

\textbf{Software-hardware co-design} (~\S\ref{sec:pipeline}).
We enable a heterogeneous CPU-GPU-SSD architecture through software-hardware co-design in \sysname. The array of NVMe SSDs provides high I/O bandwidth, while GPUs offer strong parallelism to avoid CPU computation burden during foreground search including level-0 routing centroids and level-2 vector partitions, and background clustering-based local rebuilds.

\subsection{Index Structure}
\label{sec:layout}
\sysname organizes a bottom-up three-level clustering layout across GPUs, host memory, and NVMe SSDs, supporting buffered updates and cascaded adjustments.

\subsubsection{Three-level layout}
\label{sec:three-level layout}
\begin{figure}[!t]
	\centering
	\includegraphics[width=0.9\columnwidth]{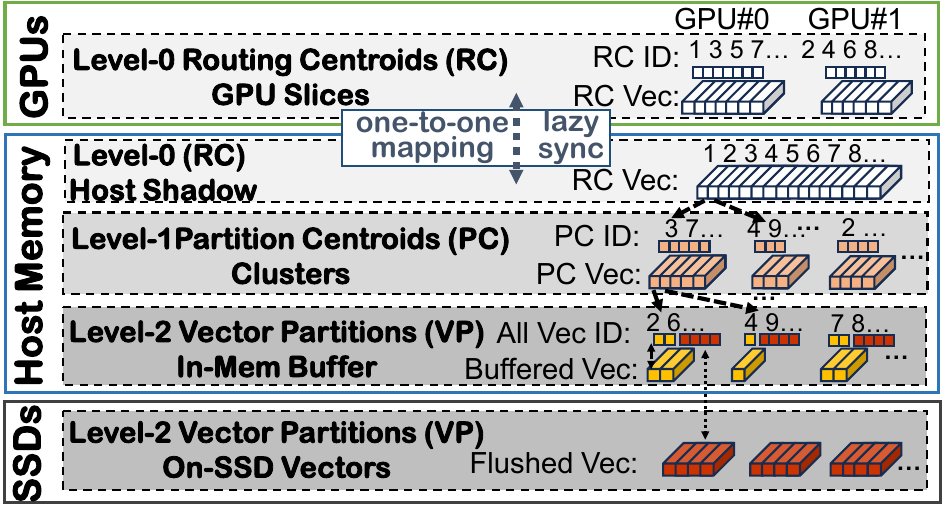}
	 \vspace{-3mm}
	\caption{Three-level storage layout of \sysname.}
	\label{fig:des6-layout}
	\vspace{-6mm}
\end{figure}

As shown in \autoref{fig:des6-layout}, \sysname organizes a three-level layout, we describe each level from bottom to top in detail below.

At \textbf{Level-2}, vectors are clustered into disjoint vector partitions (VP), where each maintains its own in-memory buffer and on-SSD continuous pages (e.g., 4 pages, 16 KB, can store 128 128-byte SIFT vectors).
Since clusters are independently updated and different clusters' pages are separated and non-contiguous on SSDs, we use per-cluster buffers rather than a shared buffer~\cite{FreshDiskANN,RocksDBExperience} to avoid inter-cluster concurrency contention and keep buffered vectors contiguous for cache-friendly access during search.
IDs of both buffered and on-SSD vectors are kept in host memory for fast deletion (\S\ref{sec:operation paths}), but stored separately in two continuous arrays to split buffered and on-SSD data so that IDs of buffered vectors can be directly deleted and IDs of on-SSD vectors can be marked for deletion with lazy garbage collection during flushes (\S\ref{sec:Cascaded Adjustment}).

At \textbf{Level-1}, partition centroids (PC) of level-2 vector partitions are clustered in memory, where each level-1 entry (i.e., a partition centroid) routes a level-2 vector partition.

At the top, \textbf{Level-0} further consists of routing centroids (RC) of clusters in level-1 partition centroids.
\sysname stores a contiguous host-side level-0 shadow with a one-to-one mapping to GPU-resident routing centroids, whose slices are distributed across GPUs for acceleration during level-0 search.
Lazy synchronization first applies cascaded level-0 centroid
adjustments to the host shadow (\S\ref{sec:Cascaded Adjustment}), then propagates them to GPU
memory after the current batched search kernel (\S\ref{sec:packing level-0}) completes and
before the next begins.
Thus, each GPU kernel uses routing centroid values that may briefly
lag the latest host-side adjustments with nearly no measurable impact on the recall.

\subsubsection{Operation paths}\label{sec:operation paths}
\begin{figure*}[!t]
	\centering
	\begin{subfigure}[t]{0.36\textwidth}
		\includegraphics[width=\linewidth]{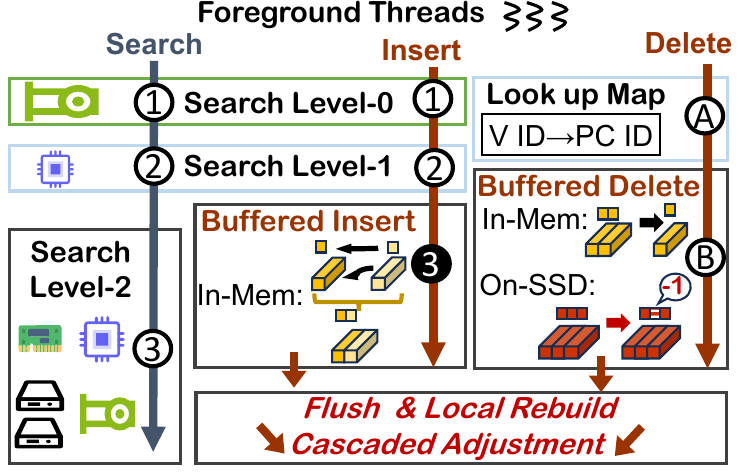}
				\vspace{-6mm}
		\caption{\centering\textbf{Foreground operation paths.}}
		\label{fig:des1-12}
		\vspace{-4mm}
	\end{subfigure}
	\centering
    	\begin{subfigure}[t]{0.62\textwidth}
		\includegraphics[width=\linewidth]{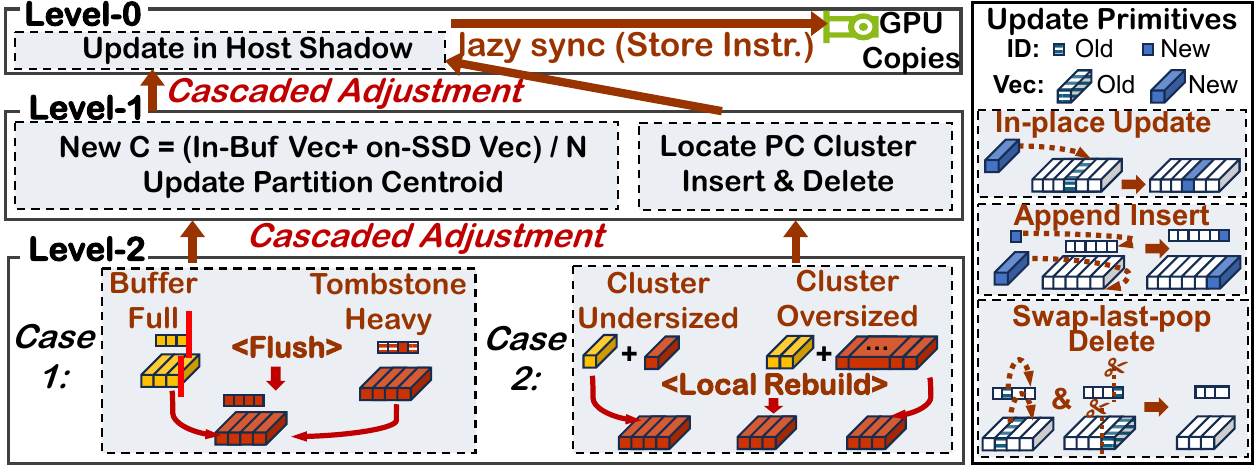}
				\vspace{-6mm}
		\caption{\centering\textbf{Flush, local rebuild, and cascaded adjustment.}}
		\label{fig:des1-13}
		\vspace{-4mm}
	\end{subfigure}
	\caption{(a) shows foreground operations, especially buffered update (i.e., insert and delete). (b) shows flush and local rebuild at level-2, which can trigger cascaded adjustment of centroids to higher levels (i.e., level-1 and level-0) with basic update primitives.}
	\label{fig:des1}
	\vspace{-3mm}
\end{figure*}

Based on the three-level layout described above, \autoref{fig:des1-12} further shows the foreground operation paths, including search, insert, and delete.

\textbf{Search} first \circleOne[0.8em]{1} searches level-0 on all flat GPU-resident slices to get candidate partition centroid clusters, and \circleOne[0.8em]{2} searches level-1 to find several nearest candidate level-2 vector partitions. Finally, it \circleOne[0.8em]{3} scans both buffered and on-SSD vectors of candidate level-2 VP to get the final top-$k$.

\textbf{Insert} also locates the target vector partition by searching level-0 and level-1 (\circleOne[0.8em]{1} and \circleOne[0.8em]{2}), but only needs the nearest one for update.
After locating the target VP, \sysname \circleNum[0.9em]{3} appends the new vector to its buffer and records the vector's ID in memory, and further records the ID of the partition that receives this new vector in a hash-based map.

\textbf{Delete} first \circleOne[0.6em]{A} uses the map structure to find the vector partition ID for the vector to be deleted. Then, \sysname \circleOne[0.6em]{B} conducts the deletion. If the vector is still in the buffer, it is removed by swap-last-pop (i.e., swapping it with the last entry and popping) to keep the layout contiguous. Otherwise, for an on-SSD vector, \sysname only marks its ID as $-1$, filters it during search, and performs lazy garbage collection during flushes, avoiding frequent SSD accesses on the foreground deletion path as well.

\subsubsection{Flush, local rebuild, and cascaded adjustment}
\label{sec:Cascaded Adjustment}
Following the above paths, foreground insert and delete can trigger flush and rebuild operations when certain conditions are met, and cause cascaded adjustments at higher levels.

In ~\autoref{fig:des1-13}, case 1 shows the level-2 \textbf{flush}. When a buffer becomes full (e.g., buffering 3 vectors reaches the budget in the figure), or its on-SSD part contains too many invalid items (e.g., 3 IDs are $-1$), a flush is triggered. It merges buffered vectors and the remaining valid on-SSD vectors, and writes them back to the same SSD region as continuous pages. The buffer is then cleared and the in-memory ID records are refreshed. Case 2 shows the background \textbf{local rebuild}, which is triggered when the vector partition becomes imbalanced following SPFresh~\cite{SPFresh} (i.e., it is undersized or oversized).
In contrast, \sysname directly performs reclustering on the imbalanced vector partition and its nearby partitions, regenerating balanced partitions instead of SPFresh's per-vector conditional reassignments.

After flush or local rebuild, \sysname{} applies
\textbf{cascaded adjustment} to propagate refreshed centroids to upper levels.
A flush summarizes multiple but bounded level-2 updates, up to the sum of buffer
capacity and tombstone threshold, into one centroid update over up-to-date members. This allows
upper-level centroids to reflect distribution changes in a timely manner while
avoiding frequent I/Os from refreshing centroids after every
foreground update. When local rebuild creates or removes vector partitions,
\sysname{} inserts new centroids into, or deletes old centroids from,
the nearest level-1 clusters. These changes are further propagated to
the host shadow with modifications to routing centroids values and lazily synchronized to GPUs by low-latency store instructions from CPUs~\cite{Neos,FlowANN}.

All upper-level changes use the same update primitives as level-2 buffer, as shown on the right of~\autoref{fig:des1-13}: in-place update for modifying centroids, append for inserting
centroids, and swap-last-pop for deleting centroids. This keeps centroid arrays
densely packed and contiguous, making them friendly to SIMD and GPU-based
distance computing.

\subsection{Concurrency Control}
\label{sec:concurrency}

\subsubsection{Decoupled locks}To maximize concurrency, we adopt decoupled locks on the hierarchical clustering structure of \sysname, where all granularities (i.e., G0, G1, G2, corresponding to partition centroid cluster,  vector partition, and vector) have independent control components (\autoref{tab:concurrency-objects}).

\begin{table}[h]
\centering
\vspace{-2mm}
\caption{Concurrency-control components in \sysname.}
\vspace{-3mm}
\label{tab:concurrency-objects}
\footnotesize
\setlength{\tabcolsep}{0.5pt}
\renewcommand{\arraystretch}{1.1}
\resizebox{\linewidth}{!}{
\begin{tabular}{@{}
>{\centering\arraybackslash}m{0.29\linewidth}
>{\centering\arraybackslash}m{0.44\linewidth}
>{\centering\arraybackslash}m{0.27\linewidth}
@{}}
\toprule
\textbf{Granularity} & \textbf{Protected Object} & \textbf{Components} \\
\midrule

\begin{tabular}[c]{@{}c@{}}
\textbf{Partition Centroid}\\
\textbf{Cluster}
\end{tabular}
&
\begin{tabular}[c]{@{}c@{}}
Level-0 routing centroid\\
Level-1 PC cluster
\end{tabular}
&
\begin{tabular}[c]{@{}c@{}}
G0 Per-cluster\\
Lock
\end{tabular}
\\

\midrule

\begin{tabular}[c]{@{}c@{}}
\textbf{Vector}\\
\textbf{Partition}
\end{tabular}
&
\begin{tabular}[c]{@{}c@{}}
Level-2 vector partition\\
Parent PC ID
\end{tabular}
&
\begin{tabular}[c]{@{}c@{}}
G1 Per-cluster\\
Lock \& Version
\end{tabular}
\\

\midrule

\begin{tabular}[c]{@{}c@{}}
\textbf{Vector}
\end{tabular}
&
\begin{tabular}[c]{@{}c@{}}
Parent vector partition ID
\end{tabular}
&
\begin{tabular}[c]{@{}c@{}}
G2 Per-slot Lock
\end{tabular}
\\

\bottomrule
\end{tabular}
}
\vspace{-3mm}
\end{table}
Across hierarchical levels, \sysname{} binds each centroid with the lower-level cluster member set it represents as one atomic concurrency-control unit.
Specifically, a level-2 vector partition and
its corresponding level-1 centroid are protected together as a vector
partition. Similarly, a level-1 partition centroid cluster and its corresponding
level-0 routing centroid are protected together as a unit. Each cluster
entry has its own lock for atomic operations. To address online rebuilds at level-2, each vector partition has an additional version record for consistency validation.

In addition, as mentioned in \S\ref{sec:layout}, \sysname{} maintains a parent vector partition ID for each vector to
support ID-based deletion and remapping. To protect these ID
mappings, but avoid expensive per-vector locks, we use
slot-based locks: each vector ID is hashed to a lock slot by ID
\texttt{mod} the count of slots. This reduces the lock memory footprint from
vector scale to slot scale (e.g., from 1B vector locks to
0.1M slot locks), while preserving concurrency.

\subsubsection{Basic principles} Relying on the above components, \sysname{} follows these three principles.

\noindent\textbf{1. Record-level concurrency consistency.} Designed for approximate top-$k$ results~\cite{SPFresh,OdinANN}, \sysname{} ensures a record-level immediate visibility~\cite{Manu,Milvus}. Once an \textit{insert} or \textit{delete} completes, its effect is visible
to subsequent searches: inserted vectors are included for
retrieval, and deleted vectors are excluded from results.

\noindent\textbf{2. Decoupling with version validation.}
To avoid unnecessary blocking (e.g., intra-cluster members should not block each other just because they belong to the same parent cluster),
\sysname{} locks only units used in the current stage and releases them
before switching to the other granularity in the next stage. Since level-2 clusters may change between stages (e.g., some of the candidate clusters returned by level-1 search might have been rebuilt), the level-2 related stage validates recorded versions before access. On mismatch, it can fall back to the previous stage or ignore invalid items.

\noindent\textbf{3. Ordered locks.}
To prevent deadlocks, operations that need multiple locks acquire them in a
deterministic order: higher-level granularity before lower-level granularity (e.g., partition centroid cluster before vector partition), and
units at the same granularity in ascending ID order (i.e., acquiring locks for clusters and slots after sorting by IDs).

\begin{figure}[!t]
	\centering
	\includegraphics[width=0.96\columnwidth]{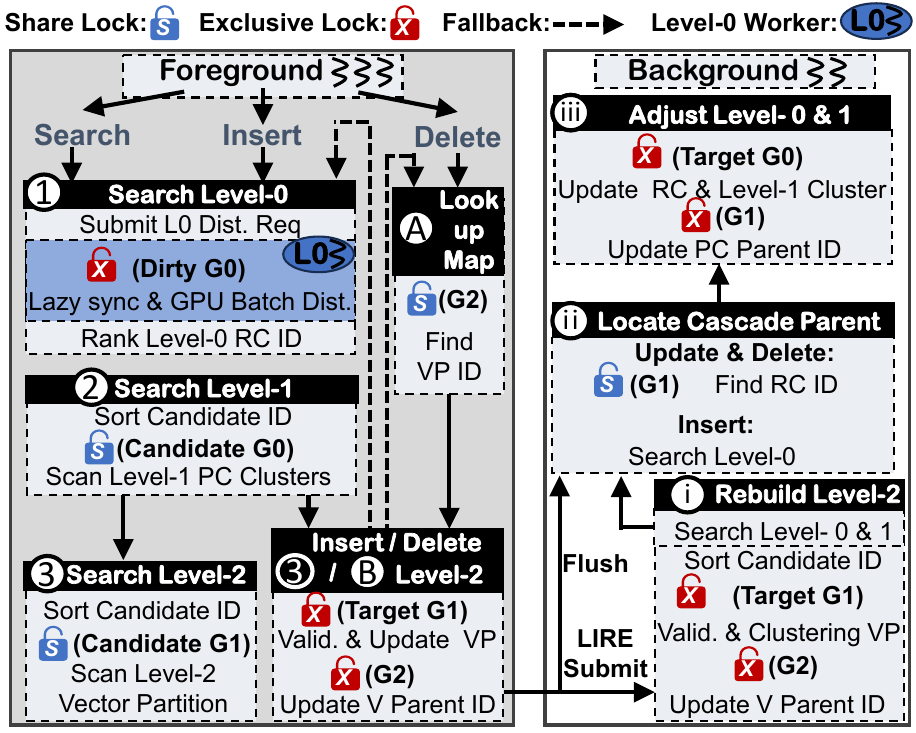}
	 \vspace{-4mm}
	\caption{Locking actions in the workflow of \sysname.}
	\label{fig:des7-control}
	\vspace{-6mm}
\end{figure}
\subsubsection{Locking mechanisms}
\label{sec:locking-mechanisms}
\sysname{} applies locks only to objects accessed by each stage, as shown in~\autoref{fig:des7-control}.

For \circleOne[0.8em]{1} search level-0,
foreground searches and insertions submit requests to level-0 workers, where each GPU has a dedicated worker. The
worker lazily synchronizes dirty level-0 centroids updated under exclusive G0 per-cluster locks,
and then performs batched GPU computation. Foreground threads only
receive distances, then rank routing centroids (RC) to select the candidate level-1 partition centroid clusters.

For \textbf{search}, \sysname{} first performs level-0 \circleOne[0.8em]{1}.
It then \circleOne[0.8em]{2} acquires shared G0 locks on candidate partition centroid clusters,
following \textit{Principle 3}. Then, it scans candidate level-1 clusters to get candidate vector partitions. Following \textit{Principle 2}, it releases G0 locks before \circleOne[0.8em]{3} acquiring shared G1 locks on candidate vector partitions, validating their versions and scanning valid vectors for final top-$k$ results. Version mismatches are handled by skipping invalid vector partitions, which keeps stable recall as evaluations of \S\ref{sec:evaluation} show, but without unnecessary retries.

For \textbf{insert}, it shares \circleOne[0.8em]{1} and \circleOne[0.8em]{2} with search to locate the nearest target level-2 vector partition to insert the new vector. It then \circleNum[0.9em]{3} acquires an exclusive G1 lock on the target vector partition (VP), validates it, and inserts the vector into it.
After that, it updates the vector-to-partition mapping under an exclusive G2 slot lock on the vector ID. If the version validation fails because of background rebuilds, it retries \circleOne[0.8em]{1} and \circleOne[0.8em]{2} to find the new target VP again.

For \textbf{delete}, it first \circleOne[0.7em]{A} reads the parent vector partition ID under a shared G2 slot lock and immediately releases it. It then acquires \circleOne[0.7em]{B}  an exclusive G1 lock on the vector partition to validate whether it still contains the vector to be deleted. If true, it deletes the vector and removes the vector-to-partition mapping under an exclusive G2 slot lock.
If the validation fails due to reposition from rebuilds, it retries reading the newest parent partition ID (\circleOne[0.7em]{A}) and tries to delete it again.

During the background \textbf{rebuild}, if a level-2 vector partition is undersized or oversized, \sysname \circleOne[0.8em]{i} performs reclustering. By \circleOne[0.8em]{1} and \circleOne[0.8em]{2}, it locates nearby vector partitions to be reclustered together and then acquires exclusive G1 locks on nearby partitions.
Valid vector partitions (VP) are then reclustered together, where their member vectors' partition IDs are remapped under G2 locks.

Further, rebuilds and flushes trigger \textbf{cascaded
adjustment} to keep upper-level centroids valid and representative. In
\circleOne[0.8em]{ii}, for updates from flushes and deletions from
reclustering, \sysname{} reads the corresponding level-1 cluster under
shared G1 locks and releases locks after lookups. For inserting a new
centroid, it locates the target partition centroid cluster through
\circleOne[0.8em]{1}. Then, in \circleOne[0.8em]{iii}, it acquires exclusive
G0 lock to update the partition centroid cluster, and modifies partition centroid
parent (i.e., routing centroid) ID under exclusive G1 lock.

 \begin{figure}[!t]
	\centering
	\begin{subfigure}[t]{0.208\textwidth}
		\includegraphics[width=\linewidth]{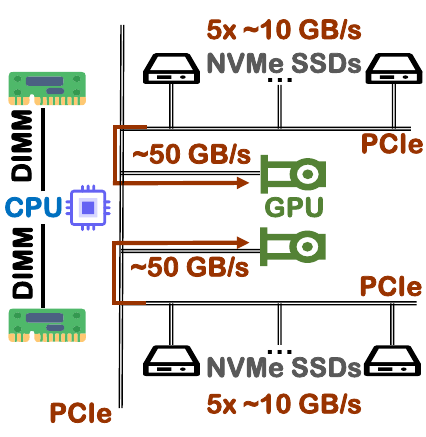}
				\vspace{-6mm}
		\caption{\centering\textbf{High-bandwidth interconnect platform.}}
		\label{fig:des3-0}
		\vspace{-3mm}
	\end{subfigure}
	\centering
	\begin{subfigure}[t]{0.23\textwidth}
		\includegraphics[width=\linewidth]{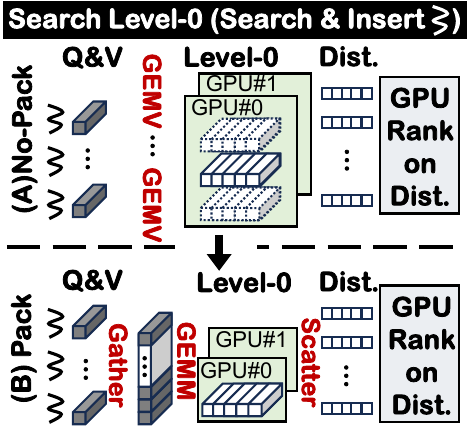}
				\vspace{-6mm}
		\caption{\centering\textbf{Packing multiple search and insert queries at level-0.}}
		\label{fig:des3-1}
		\vspace{-3mm}
	\end{subfigure}
    	\centering
	\caption{ (a) shows our hardware platform with high-bandwidth interconnects. In (b), (A) corresponds to naive no-packing and (B) shows the packing implementation.}
	\label{fig:des3}
	\vspace{-4mm}
\end{figure}

\subsection{Software-Hardware Workflow}
\label{sec:pipeline}
As shown in~\autoref{fig:des3-0}, to avoid the bandwidth bottleneck of host memory~\cite{Helmsman} and support freshness requirements,
\sysname{} runs on a CPU-GPU-SSD architecture with high-bandwidth interconnects that allow direct GPU-SSD I/O with $\sim$50 GB/s between each GPU and 5 NVMe SSDs~\cite{NVIDIAGDS,GoFS,NipstensorGDS,GeminiFS}. This architecture provides high theoretical
throughput for concurrent search and update, but also requires careful software-hardware co-design to avoid bottlenecks and underutilization. We identify three key stages that require customized optimizations: search level-0 on GPUs, search level-2 across CPUs, GPUs, and SSDs, and rebuilds on GPUs.

\subsubsection{Batched Distance Computation for Level-0.}
\label{sec:packing level-0}
As shown in~\autoref{fig:des3-1}, a naive implementation launches a level-0 scan
for each search or insert independently. It performs GEMV distance computation~\cite{FaissWiki}
followed by GPU ranking (i.e., radix-based selection~\cite{airtopk,cub}), over all
level-0 vectors; both stages are bandwidth-bound. This becomes severe at large
scale: for 1B-scale SIFT~\cite{SIFT}, level-0 can contain
$10^5$--$2\times10^5$ 128-dimensional FP16 vectors in experiments. GEMV alone reads
24--50~MB per request, requiring 1.4-3~TB/s of bandwidth at the target
throughput (i.e., 60K aggregated from 30K search and 30K update QPS~\cite{Helmsman}). Moreover,
radix-style ranking repeatedly reads and writes the distance/id array; four
read-write passes (32-bit distance / 8 bits per radix pass) add 6--13~MB per
request. These together can demand $\sim$1.8--3.8~TB/s that can exhaust the actual GPU memory bandwidth (e.g., two GPUs provide 2-2.6~TB/s in our platform as \S\ref{sec:setup}).

To address this issue, \sysname{} packs concurrent level-0 distance
computing into continuous dynamic batched GPU operations~\cite{Orca} across all GPUs, which allows for better utilization of GPU resources and reduces memory bandwidth pressure. As shown in (B) of
\autoref{fig:des3-1}, on each GPU, a worker gathers queries from search and
insert workers, forms a dense query batch, and computes query-RC (routing centroid) distances
with GEMM. It then scatters each query's distance vector back to its foreground
worker, which performs GPU ranking to select candidate partition centroid clusters.
Packing isolated GEMV into batched GEMM reuses routing centroid reads across queries,
improves arithmetic intensity, and significantly reduces per-request bandwidth
demand.

\subsubsection{Parallelizing in-buffer and on-SSD parts during level-2 search.}
\begin{figure}[!t]
    	\centering
	\begin{subfigure}[t]{0.2747\textwidth}
		\includegraphics[width=\linewidth]{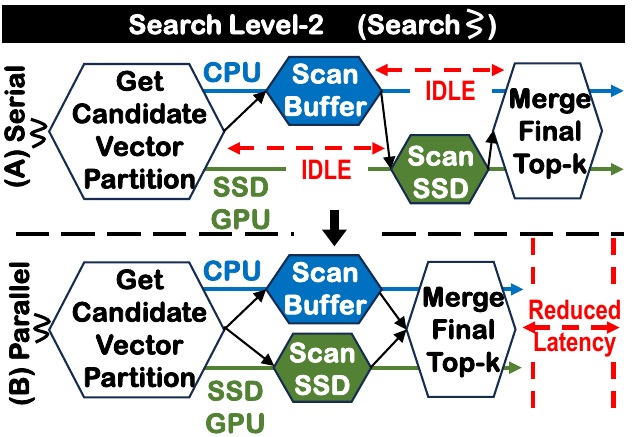}
						\vspace{-6mm}
		\caption{\centering\textbf{Parallel in-buf search on CPU and on-SSD search on direct SSD-to-GPU.}}
		\label{fig:des3-2}
		\vspace{-4mm}
	\end{subfigure}
    	\centering
	\begin{subfigure}[t]{0.1835\textwidth}
		\includegraphics[width=\linewidth]{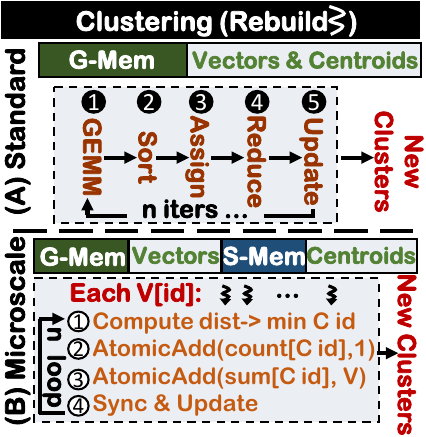}
						\vspace{-6mm}
		\caption{\centering\textbf{Fuse kernels for micro-scale reclustering.}}
		\label{fig:des3-3}
		\vspace{-4mm}
	\end{subfigure}
	\caption{ (a) shows parallel search on vector partitions. (b) shows fused clustering, where (A) and (B) correspond to a straightforward and the optimized solution, respectively.}
	\label{fig:des3-new}
	\vspace{-6mm}
\end{figure}

In~\autoref{fig:des3-2}, each selected level-2 vector partition contains two
parts: recently updated vectors in the in-memory buffer and flushed vectors on
SSDs. A naive implementation scans them sequentially, which leaves CPUs idle
during on-SSD search or leaves SSDs and GPUs idle during buffer scanning.
\sysname{} instead parallelizes these two parts, as shown in~\autoref{fig:des3-2}(B). The foreground worker scans the
small in-memory buffer on the CPU, while the on-SSD vectors are transferred to GPU
directly and scanned for distances, and then the distances are moved to CPUs. After both paths finish, their
candidates are merged by distances to get the final top-$k$ on foreground CPU workers. This overlap avoids idle hardware, thereby reducing search latency.

\subsubsection{Fusing kernels for micro-scale clustering.}
As shown in~\autoref{fig:des3-3}, local rebuilds in \sysname{} are micro-scale:
each rebuild reclusters at most 64 nearby vector partitions, usually involving thousands of vectors. At this scale, standard GPU KMeans backends~\cite{FAISSKMeansImpl,cuVSKMeansImpl} are
inefficient. Their Lloyd iterations commonly invoke multiple GPU kernels, including
GEMM-based distance computation, sorting, assignment, and reduction for new centroids, where kernel launches and intermediate global memory (i.e., G-Mem)
traffic can result in a rebuild time of up to 10-30 ms, reducing GPU efficiency.

\sysname{} therefore designs a fused micro-scale clustering kernel as shown in (B) of~\autoref{fig:des3-3}.
Because the centroids are small (e.g., 64$\times$128-dimensional FP32 vectors take only 32~KB), they can be cached in GPU shared memory (i.e., S-Mem, hundreds of KB per streaming multiprocessor). We use a design in which each GPU thread processes one vector, computes distances to all centroids, finds the nearest cluster, and immediately updates the cluster's count and vector sum through atomic accumulation.
These statistics are then used to compute centroids
for the next Lloyd iteration. By fusing assignment with statistics accumulation,
\sysname{} avoids materializing and regrouping assignments, reduces global-memory
traffic, and removes several kernel launches from the CPU side, achieving <5ms micro-scale clustering.

\begin{figure*}[t]
    \centering
    \begin{subfigure}[t]{\textwidth}
        \centering
        \includegraphics[width=\linewidth]{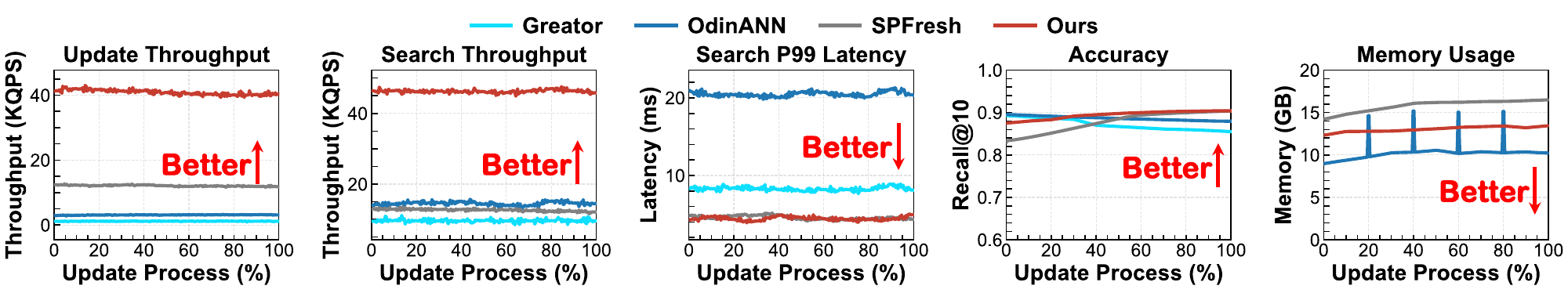}
        \vspace{-6.5mm}
        \caption{\textbf{Overall performance on \textit{Workload A} (i.e., 0.1B-scale SIFT with a uniform distribution).}}
        \label{fig:eva-11}
                \vspace{-1mm}
    \end{subfigure}
    \hfill
    \begin{subfigure}[t]{\textwidth}
        \centering
        \includegraphics[width=\linewidth]{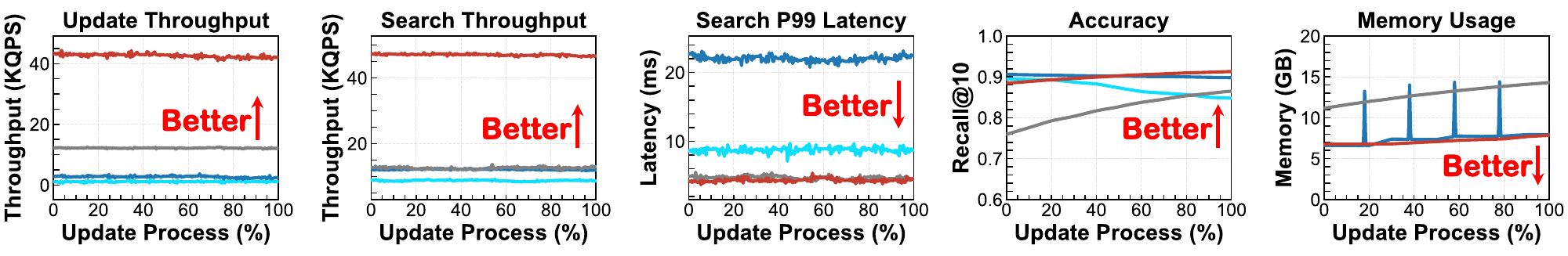}
        \vspace{-6.5mm}
        \caption{\textbf{Overall performance on \textit{Workload B} (i.e., 0.1B-scale SPACEV with a skewed distribution).}}
        \label{fig:eva-12}
    \end{subfigure}
    \hfill
    \begin{subfigure}[t]{\textwidth}
        \centering
        \includegraphics[width=\linewidth]{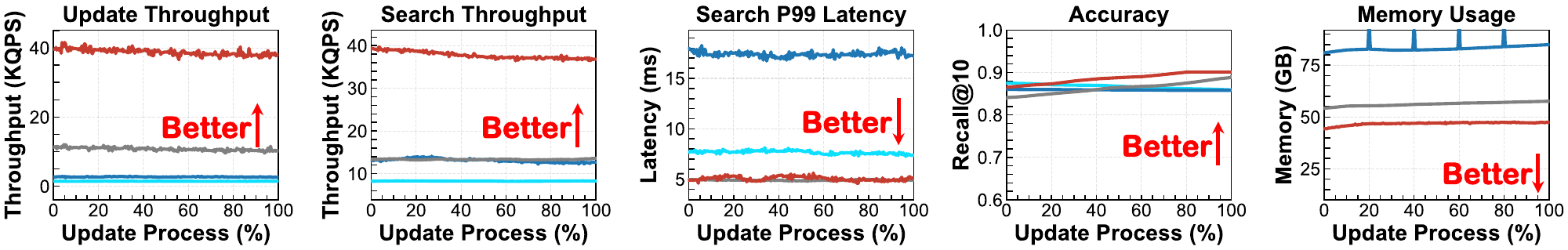}
        \vspace{-6.5mm}
        \caption{\textbf{Overall performance on \textit{Workload C} (i.e., scaling the comparison on the 1B-scale SIFT).}}
        \label{fig:eva-13}
    \end{subfigure}
    \vspace{-5mm}
    \caption{Performance comparisons about update throughput, search throughput, tail latency, accuracy, and memory usage.}
    \label{fig:eva-1}
    \vspace{-2mm}
\end{figure*}
\section{Evaluation}
\label{sec:evaluation}
We conduct evaluations for the following questions:
\begin{itemize}
    \item How does \sysname{} compare with advanced SSD-based systems by performance under standard workloads? (\S\ref{sec:overall-performance})
    \item  What are the sensitive configurations of \sysname{}? And how do \sysname{}'s techniques and extra hardware contribute to its performance improvements? (\S\ref{sec:Ablation})
    \item What is the hardware efficiency of \sysname{} in terms of device price, energy, and storage? (\S\ref{sec:scalability-cost})
    \item How does \sysname{} compare with GPU-based real-time ANNS and perform under higher dimensional datasets and production workloads? (\S\ref{sec:gpu-highdim-production})
\end{itemize}

\vspace{-3mm}
\subsection{Experimental Setup}
\label{sec:setup}
\noindent\textbf{Platform.} For evaluations of run-time performance, we use the same platform as our production services' deployment settings, which comes with an EPYC 64-core Genoa CPU, 12-channel DDR5 memory, 10$\times$ 2TB PCIe-Gen5 NVMe TLC SSDs, and 2 NVIDIA RTX Pro 5000 Blackwell GPUs.

\noindent\textbf{Workloads.} Following previous work~\cite{SPFresh,OdinANN}, we use three widely used standard dynamic workloads, unless otherwise stated.
\begin{itemize}[leftmargin=*, nosep]
    \item \textit{Workload A} simulates the scenario, where all vectors are updated (i.e., deleted and re-inserted with new embeddings) on a 100M subset of SIFT1B~\cite{SIFT}. It consists of 128-dimensional \textsf{uint8} vectors and its similarity is measured using Euclidean distance. We extract two disjoint 100M vectors from SIFT1B, where one is used as the initial dataset and the other is used for updates. Each update is performed by deleting a vector from inserted vectors and inserting a new vector from the update dataset.

    \item \textit{Workload B} has the same setting as Workload A but with 100 million vectors of SPACEV1B~\cite{SPACEV}, which consists of 100-dimensional \textsf{int8} vectors. Different from the uniform distribution of SIFT, SPACEV has a skewed distribution~\cite{SPFresh}.

    \item \textit{Workload C} scales up the experiment to 1 billion vectors, where the initial dataset is 900M vectors in SIFT1B and the update dataset is the remaining 100M vectors.
\end{itemize}

\noindent\textbf{Comparisons.}
We compare \sysname{} with state-of-the-art SSD-based ANNS systems that support updates: Greator~\cite{TopologyAwareLocalizedUpdate}, OdinANN~\cite{OdinANN} and SPFresh~\cite{SPFresh}.
The parameters are kept at optimally default values recommended, except that Greator's and OdinANN's graph-based beam width is set to 16 for better bandwidth utilization on multi-SSD setups~\cite{Helmsman}.
We do not include FreshDiskANN~\cite{FreshDiskANN} in comparisons, as it unacceptably incurs prohibitive memory consumption--up to 5-6$\times$ the dataset size and has been proven to perform worse (i.e., maximally 3K QPS updates and 5K QPS searches) than others.
Since SPFresh's open-source implementation does not support multi-SSD layout, we adapt it to support multi-SSD operation using the same storage interface as \sysname{}. Specifically, each posting list is stored in one device's contiguous pages and extended with a device identifier to route SPDK~\cite{SPDK} requests across SSDs. Graph-based SVFusion~\cite{SVFusion} is included as the state-of-the-art GPU-CPU comparison.
\begin{table}[h]
\centering
\vspace{-2.5mm}
\caption{Thread allocation. FG: foreground; BG: background.}
\label{tab:thread-counts}
\vspace{-3mm}
\resizebox{0.9\linewidth}{!}{%
\begin{tabular}{c|c|c|c|c}
\hline
& \textbf{Greator} & \textbf{OdinANN} & \textbf{SPFresh} & \textbf{Ours} \\
\hline
\textbf{Update} & 25  & 25  & FG:32 \& BG:16 & FG:16 \& BG:2 \\
\textbf{Search} & 39   & 39  & 16 & 46 \\
\hline
\end{tabular}
}
\vspace{-1.5mm}
\end{table}

\noindent\textbf{Thread allocation.} As shown in \autoref{tab:thread-counts},
we allocate threads to make the update-to-search throughput ratio $\sim$1:1 for each system, which is a common requirement in high-freshness applications~\cite{Helmsman}. Since graph-based Greator and OdinANN show limited update throughput, we set 25 threads to maximize their update throughputs first~\cite{OdinANN} and the remaining threads are used in search. In Greator, 25 update threads first conduct insertions and deletions then switch to merge, per 1\% of the process as a batch~\cite{TopologyAwareLocalizedUpdate}.

\subsection{Overall Performance}
\label{sec:overall-performance}
\subsubsection{0.1B-scale evaluation.} As shown in~\autoref{fig:eva-1}, we first evaluate all systems with \textit{Workloads A} and \textit{B}.
On both uniform (\autoref{fig:eva-11}) and skewed (\autoref{fig:eva-12}) datasets, \sysname{} outperforms the baselines in terms of throughputs of update ($\sim$3-35$\times$) and search ($\sim$3-5$\times$). And it also shows lower tail latency, comparable or better accuracy (i.e., recall@10), and the same level memory footprint.

For update, \sysname achieves $\sim$0.4 ms average latency per update, because of buffered updates with merged I/Os. Hence, only 16 foreground threads are sufficient to achieve a high update throughput. Benefiting from the optimized rebuilds (i.e., optimized GPU-based batched reclustering),  \sysname{} needs only 2 background threads to match $\sim$40K QPS foreground updates. Greator suffers from  heavy merge operations on buffered and on-SSD graphs~\cite{OdinANN}, leading to the lowest update throughput.
While avoiding heavy merge operations, OdinANN still suffers from the high latency and concurrency limitations of locks over the graph structure and cannot achieve high-throughput updates. SPFresh achieves higher update throughput than OdinANN, but still hits bottlenecks of frequent on-SSD read-modify-write operations and heavy reassignments, achieving only $\sim$10-12K QPS updates with 32 foreground threads and 16 background threads. Consequently, few threads (i.e., 16) are left for search, leading to low search throughput.

\sysname{} can allocate more threads (i.e., 46) to search, achieving much higher search concurrency. With 10 Gen5 NVMe SSDs providing more than 100 GB/s of bandwidth and GPUs handling computation, the I/O and host-memory-bandwidth bottlenecks in the search path can be overcome. As a result, \sysname{} reaches comparable recall by searching more clusters (e.g., $\sim$200) without throughput loss. With normal-scale partitions (i.e., 1-2\% of the dataset size), \sysname can buffer up to 8 vectors per cluster, while having a stable and acceptable memory usage. While OdinANN's memory can fluctuate due to merges on deletions, SPFresh has a higher memory usage than ours due to more centroids. Since Greator has unacceptably high memory usage (i.e., $>40$ GB for the 0.1B scale and $>280$ GB for the 1B scale) due to buffered graph and global merge, it is omitted in figures of memory usage.

\subsubsection{1B-scale evaluation.} Then, we scale the experiment to 1 billion vectors with \textit{Workload C} (\autoref{fig:eva-13}). Results are similar to the 0.1B-scale evaluation. Although 1B-scale index has $\sim$10$\times$ centroids of centroids (i.e., level-0) on GPUs, which can incur heavier GPU bandwidth consumption than 0.1B-scale index, \sysname still achieves $\sim$3$\times$ higher search throughput and up to $\sim$27$\times$ higher update throughput than the baselines. Because we pack distance computation of level-0 search together with batched GEMM, much global memory traffic can be saved and both throughputs of updates and searches still reach 36-39K QPS.
Additionally, the recall is also higher than baselines.

\vspace{-3mm}

\subsection{Sensitivity and Ablation Study}
\label{sec:Ablation}
\begin{figure}[t]
    \centering
    \begin{subfigure}[t]{0.49\linewidth}
        \centering
        \includegraphics[width=\linewidth]{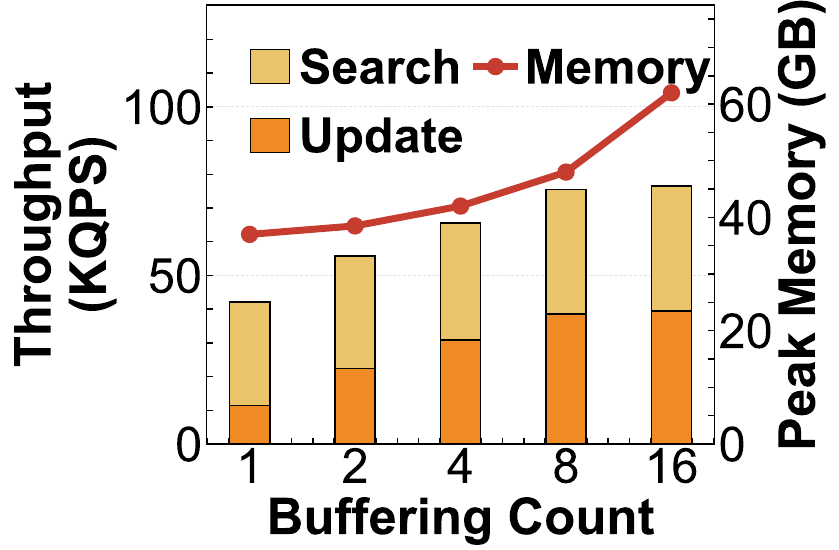}
        \vspace{-6mm}
        \caption{\centering\textbf{Throughput and memory.}}
        \label{fig:eva-21}
    \end{subfigure}
    \hfill
    \begin{subfigure}[t]{0.49\linewidth}
        \centering
        \includegraphics[width=\linewidth]{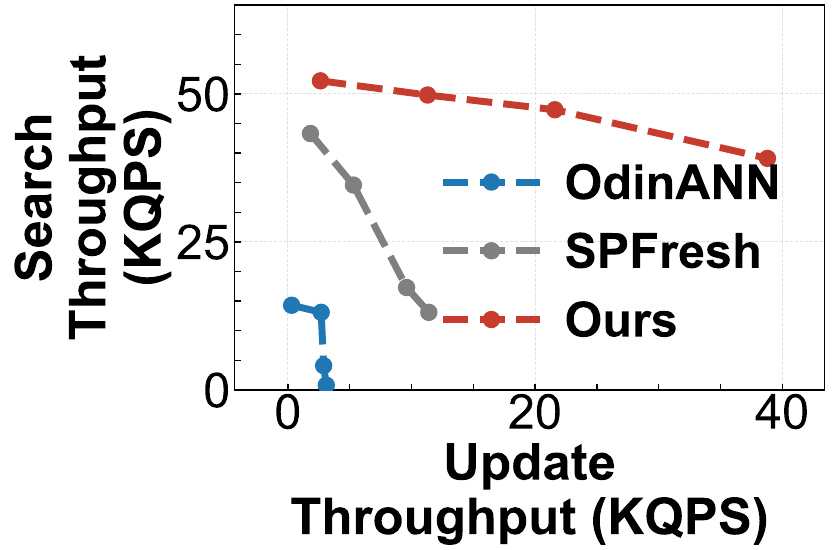}
        \vspace{-6mm}
        \caption{\centering\textbf{Update and search.}}
        \label{fig:eva-22}
    \end{subfigure}
    \vspace{-4mm}
    \caption{Resource and performance trade-offs.}
    \label{fig:eva-2}
    \vspace{-4mm}
\end{figure}

\noindent\textbf{Trade-off between throughput and memory.} With \textit{Workload C} and settings as in~\autoref{tab:thread-counts}, \autoref{fig:eva-21} shows the trade-off between throughput and memory usage with varying buffering budgets (i.e., vectors buffered and tombstones deleted per cluster). With higher memory for buffers, the update throughput can be improved by batching SSD accesses. Buffering 8 vectors can achieve a good balance between throughput and memory usage, and improvements become marginal with higher budgets (e.g., 16).

\begin{figure}[t]
    \centering
    \begin{subfigure}[t]{0.45\linewidth}
        \centering
        \includegraphics[width=\linewidth]{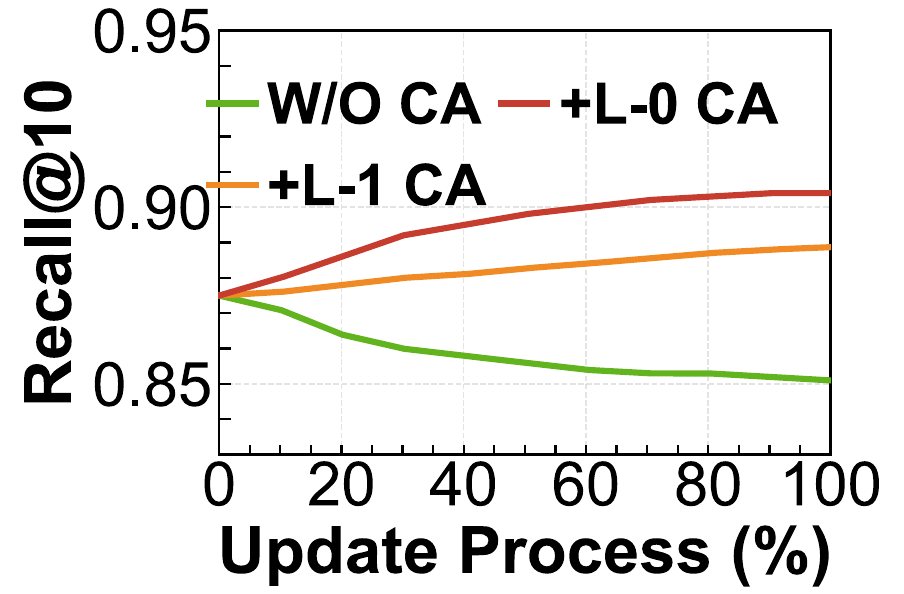}
        \vspace{-6mm}
        \caption{\centering\textbf{Cascaded adjustment.}}
        \label{fig:eva-23}
    \end{subfigure}
    \hfill
    \begin{subfigure}[t]{0.54\linewidth}
        \centering
        \includegraphics[width=\linewidth]{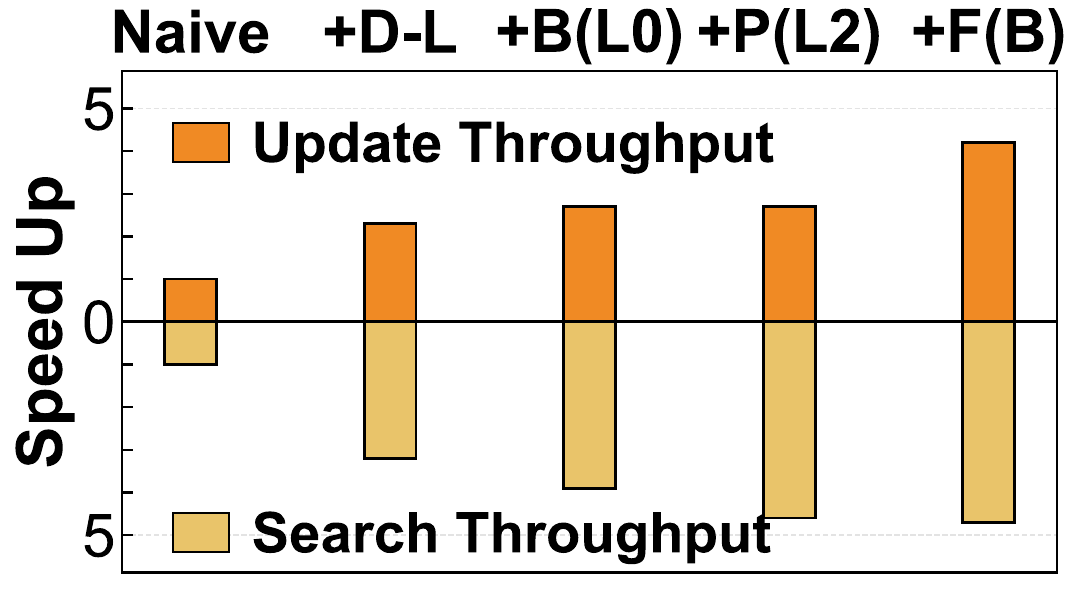}
        \vspace{-6mm}
        \caption{\centering\textbf{Performance improvements.}}
        \label{fig:eva-24}
    \end{subfigure}
    \vspace{-4mm}
    \caption{Ablation study of accuracy and performance.}
    \label{fig:eva-2-1}
    \vspace{-4mm}
\end{figure}
\noindent\textbf{Trade-off between update and search.} \autoref{fig:eva-22} shows the trade-off between update and search with varying thread allocation from search-heavy setting to update-heavy setting on \textit{Workload C}. Compared to the most advanced baselines of graph-based and partition-based methods, our \sysname can achieve both high update and search throughputs across varying ratios of update to search throughput. OdinANN suffers from the high latency and concurrency limitations of the graph structure, and thus cannot achieve high-throughput updates. SPFresh achieves higher update throughput than OdinANN, but more cores than \sysname are needed to improve update throughput, which leads to a significant drop in search throughput.

\noindent\textbf{Cascaded adjustment.} \autoref{fig:eva-23} shows the impact of cascaded adjustment on recall rate in \textit{Workload A}. Without cascaded adjustment (W/O CA), the recall rates can drop by up to 3\% in the process, even with rebuilds to improve the index quality~\cite{SPFresh}. Level-1 and Level-0 adjustments (i.e., L-1 CA and L-0 CA) can avoid the loss of efficacy and improve recall rates by up to 1.5\% and 2.9\%.

\noindent\textbf{Performance ablation.} \autoref{fig:eva-24} shows the throughput improvements by \textit{Workload C}, from software designs. Naive implementation (Naive) has a coarse-grained lock that objects from the same upper-level cluster are locked together, which can lead to high contention and low throughput. With decoupled locks (i.e., D-L) in \S\ref{sec:concurrency}, throughputs of update and search can be improved by up to 2.3$\times$ and 3.2$\times$, respectively. Batched GPU level-0 search (i.e., B(L0)) can further improve throughputs by 0.4-0.7$\times$ naive's performance. And parallelizing in-buffer and on-SSD search (i.e., P(L2)) can further improve search throughputs to 4.6$\times$. And fused background micro-scale clustering (i.e., F(B)) can further improve update throughput by $\sim$1.5$\times$, because fewer background threads can support more foreground update threads.

\begin{figure}[t]
    \centering
    \begin{subfigure}[t]{0.62\linewidth}
        \centering
        \includegraphics[width=\linewidth]{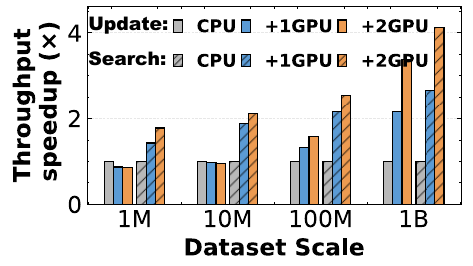}
        \vspace{-6mm}
        \caption{\centering\textbf{GPU contribution.}}
        \label{fig:eva-hw-1}
    \end{subfigure}
    \hfill
    \begin{subfigure}[t]{0.37\linewidth}
        \centering
        \includegraphics[width=\linewidth]{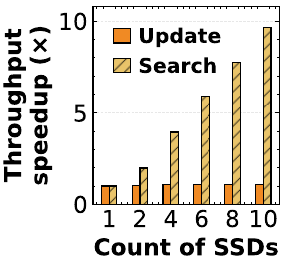}
        \vspace{-6mm}
        \caption{\centering\textbf{SSD contribution.}}
        \label{fig:eva-hw-2}
    \end{subfigure}
    \vspace{-4.5mm}
    \caption{Extra hardware contribution for \sysname.}
    \label{fig:eva-hw}
    \vspace{-4mm}
\end{figure}

\autoref{fig:eva-hw} shows the contribution of additional hardware (i.e., GPUs and more SSDs). In~\autoref{fig:eva-hw-1}, CPU means CPU-only \sysname with buffering updates and clustering-based rebuilds, while +1GPU offloads level-0 search and level-2 on-SSD search from half of 10 SSDs to a GPU. +2GPU is the fully equipped configuration with two GPUs and 10 SSDs. With different scale subsets from SIFT1B, as the scale increases, the bandwidth burden of level-0 flat search and level-2 on-SSD search becomes more significant, GPUs can bring up to 3.4$\times$ and 4.1$\times$ improvement in update and search.
In~\autoref{fig:eva-hw-2}, the contribution of SSDs is illustrated, showing how additional SSDs enhance the overall system throughput. Since update is not I/O-bound, the improvement is mainly observed in search operations. With more SSDs, \sysname can support linear scaling in search performance under concurrent update and search.

\begin{table}[h]
\vspace{-3mm}
\centering
\caption{Our hardware costs~\cite{DRAM,AMD4th,Pro5000}. TCA: total cost of acquisition; TDP: thermal design power; CAP: capacity. The chassis consists of mainboard and other components.}
\label{tab:cost}
\vspace{-3mm}
\small
\setlength{\tabcolsep}{1pt}
\renewcommand{\arraystretch}{1.1}
\resizebox{0.9\linewidth}{!}{%
\begin{tabular}{c|c|c|c|c|c|c}
\hline
 & \textbf{Chassis} & \textbf{CPU} & \textbf{GPU} & \textbf{Mem} & \textbf{SSD} & Total \\
\hline
\textbf{TCA} & 1 $\times$ \$3K  & 1 $\times$ \$4K  & 2 $\times$ \$6K & 12 $\times$ \$2K & 10 $\times$ \$0.4K & \$47K\\
\hline
\textbf{TDP} & 1 $\times$ 50W & 1 $\times$ 400W & 2 $\times$ 300W & 12 $\times$ 10W & 10 $\times$ 8W &  1250W\\
\hline
\textbf{CAP} & -- & -- & 96 GB & 768 GB & $\sim$20,000 GB & -- \\
\hline
\multicolumn{3}{c|}{\textbf{Cost Per Unit Storage}} & 125 \$/GB & 31 \$/GB & 0.2 \$/GB & --\\
\hline
\end{tabular}%
}
\vspace{-5mm}
\end{table}

\subsection{Hardware Efficiency}
\label{sec:scalability-cost}

\begin{figure}[t]
    \centering
    \begin{subfigure}[t]{\linewidth}
        \centering
        \includegraphics[width=0.7\linewidth]{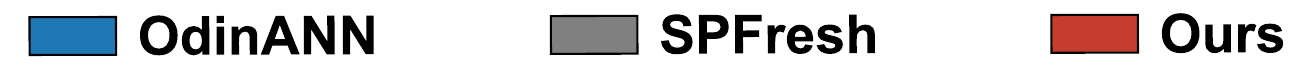}
        \vspace{-5.5mm}
        \label{fig:eva-30}
    \end{subfigure}
    \centering
    \begin{subfigure}[t]{0.31\linewidth}
        \centering
        \includegraphics[width=\linewidth]{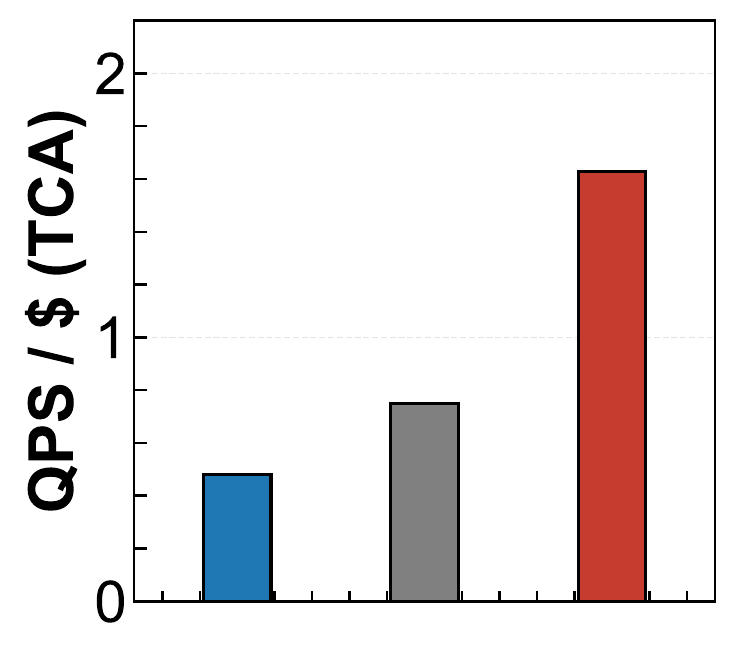}
        \vspace{-6.5mm}
        \caption{\centering\textbf{TCA efficiency.}}
        \label{fig:eva-31}
    \end{subfigure}
    \hfill
    \begin{subfigure}[t]{0.31\linewidth}
        \centering
        \includegraphics[width=\linewidth]{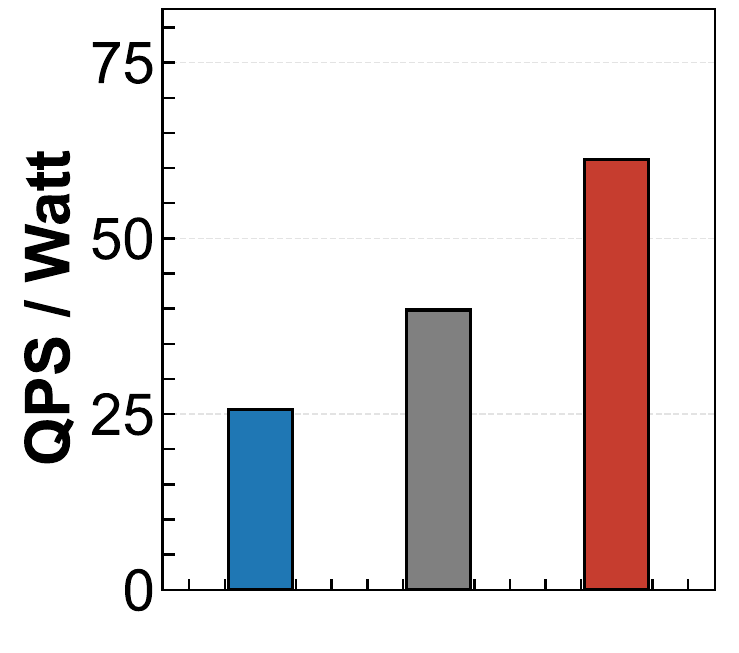}
        \vspace{-6.5mm}
        \caption{\centering\textbf{Power efficiency.}}
        \label{fig:eva-32}
    \end{subfigure}
        \begin{subfigure}[t]{0.31\linewidth}
        \centering
        \includegraphics[width=\linewidth]{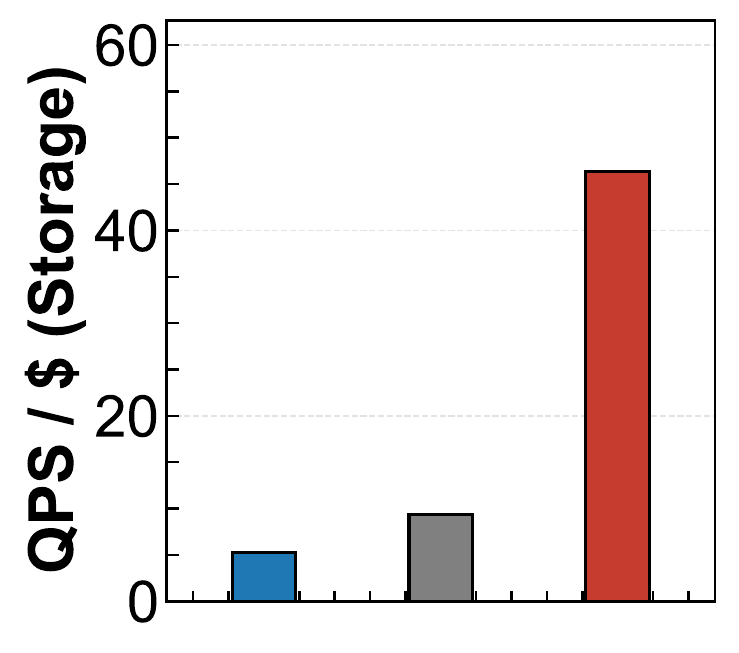}
        \vspace{-6.5mm}
        \caption{\centering\textbf{Storage efficiency.}}
        \label{fig:eva-33}
    \end{subfigure}
    \vspace{-4mm}
    \caption{Cost efficiency on device price, energy, and storage.}
    \label{fig:eva-3}
    \vspace{-4.5mm}
\end{figure}

Based on evaluations on SIFT1B, \autoref{fig:eva-3} shows the hardware efficiency of \sysname{} in terms of device price, energy, and storage footprint. The information about hardware costs in our platform is shown in~\autoref{tab:cost}. For OdinANN and SPFresh, we minimize baselines' hardware with no GPUs and fewer SSDs (i.e., 2-4 SSDs), which can reach the performance of \S\ref{sec:overall-performance} and reduce the cost to $\sim$32K and the total power to $\sim$600W, respectively.

\noindent\textbf{Efficiency of acquisition cost.}
We use aggregate throughputs of updates and searches on \textit{Workload C}  under the thread configurations in ~\autoref{tab:thread-counts} as the performance metric, and use the total cost of acquisition (TCA) from \autoref{tab:cost} as the cost metric.
As shown in ~\autoref{fig:eva-31}, although \sysname{} requires two additional GPUs and thus has about $1.5\times$ higher total device cost than OdinANN and SPFresh, it achieves $3$--$5\times$ higher aggregate throughput, leading to a $2.2$--$3.4\times$ improvement in devices' cost efficiency.

\noindent\textbf{Power efficiency.} We use the aggregate throughput as the performance metric, and use the thermal design power (TDP) as the cost metric, when evaluating power efficiency.
As shown in~\autoref{fig:eva-32}, although \sysname{} incurs about 650W higher consumption than OdinANN and SPFresh due to additional GPUs, it still achieves a $\sim$$1.5$--$2.4\times$ improvement in cost efficiency in terms of QPS per watt.

\noindent\textbf{Storage-cost efficiency.}
Following recent work~\cite{Helmsman,GustANN}, we evaluate storage-cost efficiency by combining the index footprint on each storage device with its unit cost.
For multi-SSD SPFresh, each cluster occupies 16~KB SSD space but needs only one block-mapping entry in memory.
Thus, it adds 2.5~TB SSD space while saving about 10~GB memory, making its storage cost 8--10\% higher than the single-SSD version, i.e., about 2.55K\$.
OdinANN reserves twice the SSD space for updates, leading to over 1.3~TB SSD and 85~GB memory usage, costing about 2.9K\$.

In contrast, though GPU memory is additionally introduced in \sysname, only centroid-cluster centroids are stored on GPUs, whose GPU memory footprint of the SIFT1B index is below 50~MB (<\$10 under ~\autoref{tab:cost}).
By reducing centroid counts, \sysname{} also reduces SSD and host-memory usage to about 0.4~TB and 50~GB, respectively.
The host memory mainly includes 14~GB update buffers, 3-4~GB centroids, and others for metadata such as IDs, parent mappings, and locks structure.
As a result, the storage cost of \sysname{} is $\sim$1.65K\$.
As shown in~\autoref{fig:eva-33}, this achieves $4.9$--$8.7\times$ storage-cost efficiency of baselines.

\subsection{GPU-based comparison, high-dimensional datasets, and production evaluation}
\label{sec:gpu-highdim-production}

\begin{figure}[t]
    \centering
    \begin{subfigure}[t]{0.3\linewidth}
        \centering
        \includegraphics[width=\linewidth]{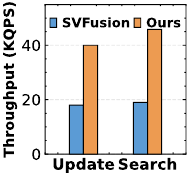}
        \vspace{-6.5mm}
        \caption{\centering\textbf{CPU-GPU.}}
        \label{fig:eva-add-1}
    \end{subfigure}
    \hfill
    \begin{subfigure}[t]{0.275\linewidth}
        \centering
        \includegraphics[width=\linewidth]{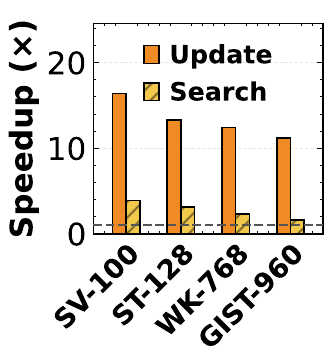}
        \vspace{-6.5mm}
        \caption{\centering\textbf{Dimensions.}}
        \label{fig:eva-add-2}
    \end{subfigure}
        \begin{subfigure}[t]{0.40\linewidth}
        \centering
        \includegraphics[width=\linewidth]{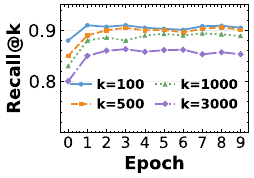}
        \vspace{-6.5mm}
        \caption{\centering\textbf{Long-term stability.}}
        \label{fig:eva-add-3}
    \end{subfigure}
    \vspace{-4mm}
    \caption{CPU-GPU, high-dim and long-term comparisons.}
    \label{fig:eva-add}
    \vspace{-4.5mm}
\end{figure}

We compare \sysname with SVFusion~\cite{SVFusion}, a state-of-the-art graph-based real-time GPU ANNS, replicating its index to maximize multi-GPU performance. We use the optimal CPU–GPU mode of SVFusion given its lack of I/O optimizations. Our system delivers approximately $2\times$ higher update and search throughput over \textit{Workload A} as shown in \autoref{fig:eva-add-1}, suggesting that partition-based methods better suit concurrent update–search workloads on GPUs.

In addition to \textit{Workloads A} and \textit{B} of SIFT and SPACEV (i.e., ST-128 and SV-100), we evaluate 768-dimensional Wikipedia (i.e., WK-768)~\cite{Wikipedia} and 960-dimensional Gist (i.e., GS-960)~\cite{GIST} against OdinANN, the state-of-the-art graph-based system, as baselines in update–search workloads. Though \autoref{fig:eva-add-2} shows diminishing gain of our system as dimensionality increases due to higher search I/O than graph-based methods, \sysname maintains over $\sim10\times$ update throughput and $\sim1.5\times$ search throughput.

We further evaluate \sysname over multiple epochs in one of our production 128-dim recommendation workloads (\autoref{fig:eva-add-3}), where each epoch corresponds to an embedding model update and a full index refresh. During the first refresh, clustering through local rebuilds improves centroid quality, yielding 3-5\% recall gains. Then, recall would remain stable across subsequent epochs.

\section{Deployment}
\begin{figure}[!t]
	\centering
	\includegraphics[width=0.98\columnwidth]{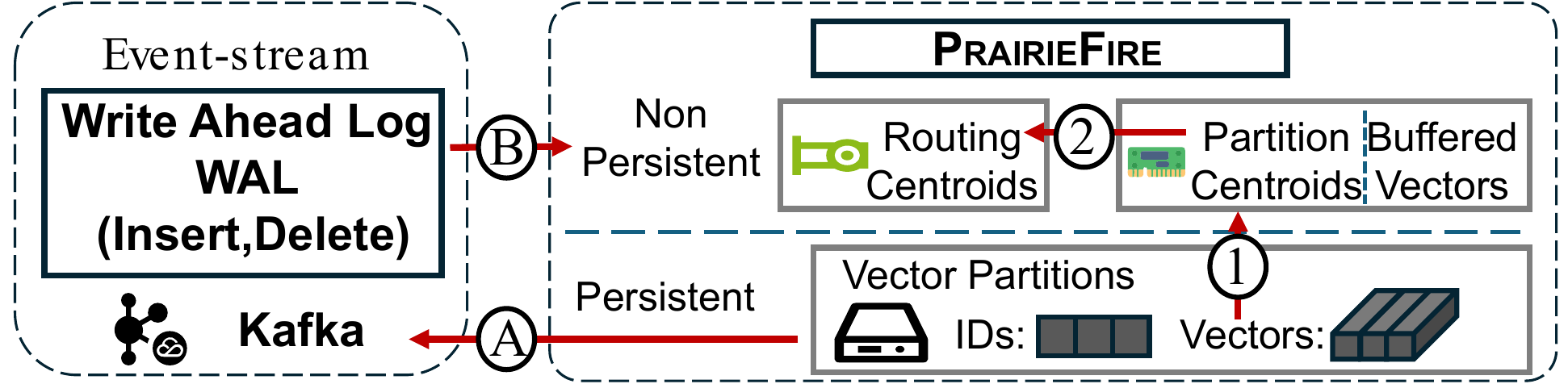}
	 \vspace{-3mm}
	\caption{{Write-ahead log (WAL) for crash recovery.}}
	\label{fig:add-wal}
	\vspace{-7mm}
\end{figure}

\noindent\textbf{Crash Recovery.} Addressing durability of crash recovery, we apply the Kafka-based write-ahead log (WAL)~\cite{Kafka} and unique incremental ID assignment mechanism with \sysname in  our industrial deployments~\cite{Manu}. This reuses the existing infrastructure, avoiding a redundant WAL implementation in \sysname. As shown in \autoref{fig:add-wal}, during crash recovery, \sysname first \circleOne[0.7em]{A} sends the vector IDs persisted on SSDs to the event-stream servers, which compare them against the complete set of valid IDs in the WAL to identify updates (i.e., inserts and deletes) that require replay. Locally, \sysname reconstructs \circleOne[0.8em]{1} partition centroids and then \circleOne[0.8em]{2} routing centroids from the persisted vector partitions. The updates requiring replay are subsequently \circleOne[0.7em]{B} dispatched from Kafka to \sysname nodes to restore the crashed index.

\noindent\textbf{Deployment status.} Addressing both high throughput and freshness requirements, \sysname has been deployed in RedNote's search, recommendation, and advertising services for two months. As an example, for one recommendation index whose embeddings are refreshed twice per hour, \sysname reduced full index rebuilds from 48 per day to one, handling embedding updates across all intervening epochs through streaming rebuilds. Replacing periodic rebuilds with streaming updates improved recommendation quality, with business metrics including exposure, click-through rate, and revenue yield increasing by 0.5--2 percentage points.

\section{Related Work}
As real-world workloads now require ANNS systems to support billions of vectors~\cite{JDPlatform,BigANN,Vexless}, in-memory solutions~\cite{FAISS,HNSW,NSG,RAIRS,ELPIS} are replaced by disk-based approaches. DiskANN~\cite{DiskANN} and Starling~\cite{Starling} are representative examples following the graph-based pattern. PipeANN~\cite{PipeANN} further optimizes the graph-based approach by pipelining the search process. These graph-based systems have higher throughput than cluster-based SPANN~\cite{SPANN} on a single SSD, because they consume less I/O bandwidth. However, as the array of modern SSDs becomes more powerful, cluster-based approaches can also achieve high throughput with low latency~\cite{SPANN,SPFresh,Helmsman,TridentANN}, becoming popular again. At the same time, freshness requirements are also increasing~\cite{FreshRec,Monolith},  FreshDiskANN, Greator, OdinANN, SPFresh, and SVFusion~\cite{FreshDiskANN,TopologyAwareLocalizedUpdate,OdinANN,SPFresh,SVFusion} are proposed to support updatable ANNS in large-scale scenarios, but their performance is limited under high update rates.
Quake~\cite{Quake} is proposed for CPU-only index maintenance and does not support concurrent update and search.
GPU-based, FPGA-based, and CXL-based ANNS~\cite{GustANN,SGIANNS,CAGRA,FusionANNS,FlashANNS,CHAVS,Falcon,CXLANNS} are also proposed to address the limitation of CPUs, but they do not support updates.

\section{Conclusion}
We argue that buffered update and cluster-based index are effective for supporting updatable ANNS in large-scale scenarios. We propose \sysname, a CPU-GPU-SSD ANNS system that supports high-throughput updates, preserving high-throughput search. Evaluations show that it can achieve the performance requirements of high-freshness applications while maintaining high cost efficiency, energy efficiency, and storage efficiency.

\bibliographystyle{ACM-Reference-Format}
\bibliography{HyperVec-Fresh}

\end{document}